\documentclass[ aip, jmp, reprint,]{revtex4}
\usepackage{amsfonts}
\usepackage{url}
\usepackage{amssymb}
\usepackage{amsmath}
\usepackage{graphicx}
\usepackage{dcolumn}
\usepackage{bm}

\begin{document}

\title[]{\textbf{Exact solution of the EM radiation-reaction problem \\
for classical finite-size and Lorentzian charged particles}}
\author{Claudio Cremaschini}
\affiliation{International School for Advanced Studies (SISSA), Trieste, Italy}
\affiliation{Consortium for Magnetofluid Dynamics, University of Trieste, Trieste, Italy}
\author{Massimo Tessarotto}
\affiliation{Department of Mathematics and Informatics, University of Trieste, Trieste,
Italy}
\affiliation{Consortium for Magnetofluid Dynamics, University of Trieste, Trieste, Italy}
\date{\today }

\begin{abstract}
An exact solution is given to the classical electromagnetic (EM)
radiation-reaction (RR) problem, originally posed by Lorentz. This refers to
the dynamics of classical non-rotating and quasi-rigid finite size particles
subject to an external prescribed EM field. A variational formulation of the
problem is presented. It is shown that a covariant representation for the EM
potential of the self-field generated by the extended charge can be uniquely
determined, consistent with the principles of classical electrodynamics and
relativity. By construction, the retarded self 4-potential does not possess
any divergence, contrary to the case of point charges. As a fundamental
consequence, based on Hamilton variational principle, an exact
representation is obtained for the relativistic equation describing the
dynamics of a finite-size charged particle (RR equation), which is shown to
be realized by a second-order delay-type ODE. Such equation is proved to
apply also to the treatment of Lorentzian particles, i.e., point-masses with
finite-size charge distributions, and to recover the usual LAD equation in a
suitable asymptotic approximation. Remarkably, the RR equation admits both
standard Lagrangian and conservative forms, expressed respectively in terms
of a non-local effective Lagrangian and a stress-energy tensor. Finally,
consistent with the Newton principle of determinacy, it is proved that the
corresponding initial-value problem admits a local existence and uniqueness
theorem, namely it defines a classical dynamical system.
\end{abstract}

\maketitle


\section{Introduction}

An unsolved theoretical problem is related to the description of the
dynamics of classical charges with the inclusion of their electromagnetic
(EM) self-fields, the so-called \emph{radiation-reaction} (RR) \emph{problem}
(Dirac \cite{Dirac1938}, Pauli \cite{Pauli1958}, Feynman \cite{Feynman1988}%
). Despite efforts spent by the scientific community in more than one
century of intensive theoretical research, an exact solution is still
missing (see related discussion in Ref.\cite{Dorigo2008a}; for a review see
Refs.\cite{Rohrlich1965,Teitel1970a,Teitel1970b,parrot1987,parrot1993}). In
this regard, of fundamental importance is the construction of the \emph{%
exact }\emph{\ (i.e., non-asymptotic)}\emph{\ relativistic equation of
motion for a classical charged particle in the presence of its EM
self-field, }also known as \emph{RR equation}. This concerns, in particular,
its treatment in the context of \emph{special relativity} (SR) and \emph{%
classical electrodynamics }(CE), namely imposing the following basic
physical requirements, hereafter referred to as \emph{SR-CE Axioms}:

\begin{enumerate}
\item[1] Axiom \#1: the Maxwell equations are fulfilled everywhere in the
flat space-time $M^{4}\mathcal{\subseteq }{\mathbb{R}}^{4}$, with metric
tensor $g_{\mu \nu }$. The Minkowski metric tensor is denoted as $\eta _{\mu
\nu }\equiv diag(1,-1,-1,-1)$. In particular the EM 4-potential $A^{\mu }$
is assumed of class $C^{k}(M^{4})$, with $k\geq 2$;

\item[2] Axiom \#2: the Hamilton variational principle holds for a suitable
functional class of variations $\left\{ f\right\} $. In particular, the
Hamilton principle must uniquely prescribe the particle world-line as a real
function $r^{\mu }(s)\in C^{k}(\mathbb{R}),$ with $k\geq 2$ for all $s\in
\mathbb{R}$. The RR equation is then determined by the corresponding
Euler-Lagrange (E-L) equations. Hence, $\left\{ f\right\} \equiv \left\{
f_{i}(s),i=1,n\right\} $ is identified with the set of real functions of
class $C^{k}(\mathbb{R}),$ with $k\geq 2$:%
\begin{equation}
\left\{ f\right\} \equiv \left\{
\begin{array}{c}
f_{i}(s):f_{i}(s)\in C^{k}(\mathbb{R}); \\
i=1,n;\emph{\ and }k\geq 2%
\end{array}%
\right\} ,  \label{FUNCTIONAL CLASS}
\end{equation}%
with functions $f_{i}(s)$ (for $i=1,n$) to be properly defined. In
particular, we shall require that the action functional is allowed to be of
the general form
\begin{equation}
S_{1}(f,\left[ f\right] )\equiv \int_{-\infty }^{+\infty }dsL_{1}\left( f(s),%
\frac{df(s)}{ds},\left[ f(s)\right] ,\left[ \frac{df(s)}{ds}\right] \right) .
\label{HAMILTON ACTION}
\end{equation}

Here $L_{1}$ denotes a non-local \emph{variational particle Lagrangian}, by
assumption defined on a \emph{finite-dimensional} phase-space, which depends
at most on first-order derivatives $\frac{df(s)}{ds},$ with $f(s)$ belonging
to the functional class $\left\{ f\right\} ,$ while $\left\{ f(s),\frac{df(s)%
}{ds}\right\} $ and $\left\{ \left[ f(s)\right] ,\left[ \frac{df(s)}{ds}%
\right] \right\} $ indicate respectively local and non-local dependencies in
terms of $f(s)$ and $\frac{df(s)}{ds};$

\item[3] Axiom \#3: the Newton determinacy principle (NDP) holds. This
implies the validity of an existence and uniqueness theorem for the
corresponding E-L equations. As a consequence, there exists necessarily a
\emph{classical dynamical system}, namely a diffeomorphism
\begin{equation}
\mathbf{x}_{0}\equiv \mathbf{x}\left( s_{0}\right) \rightarrow \mathbf{x}%
\left( s\right) ,  \label{dynsys}
\end{equation}%
with $\mathbf{x}\in I$ and $s$ representing respectively the state of a
classical particle and a suitable proper time, where $I\subseteq \mathbb{R}$
is an appropriate finite interval of the real axis;

\item[4] Axiom \#4: the Einstein causality principle (ECP) and the Galilei
inertia principle (GIP) both apply;

\item[5] Axiom \#5: the general covariance property of the theory, and in
particular the so-called manifest Lorentz covariance (MLC), i.e., the
covariance with respect to the group of special Lorentz transformations, are
satisfied.
\end{enumerate}

Manifestly, these axioms are understood as \emph{identically }fulfilled,
i.e., they must apply for arbitrary choices of both the initial conditions
for the dynamics of the charged particles and the applied external EM field.

The RR problem was first posed by Lorentz in his historical work (Lorentz,
1892 \cite{Lorentz}; see also Abraham, 1905 \cite{Abraham1905}). Traditional
approaches are based either on the RR equation due to Lorentz, Abraham and
Dirac (first presented by Dirac in 1938 \cite{Dirac1938}), nowadays
popularly known as the \emph{LAD equation}, or the equation derived from it
by Landau and Lifschitz \cite{LL} via a suitable \textquotedblleft reduction
process\textquotedblright\ , the so-called \emph{LL equation}. As recalled
elsewhere \cite{Dorigo2008a} several aspects of the RR problem - and of the
LAD and LL equations - are yet to find a satisfactory formulation/solution.\
Common feature of all previous approaches is the adoption of an asymptotic
expansion for the EM self-field (or for the corresponding EM 4-potential),
rather than of its exact representation. This, in turn, implies that such
methods allow one to determine - at most - only an asymptotic approximation
for the correct RR equation.

For contemporary science the solution of the RR problem represents a
fundamental prerequisite for the proper formulation of all relativistic
theories, both classical and quantum ones, which are based on the
description of relativistic dynamics for classical charged particles.

Since Lorentz famous paper \cite{Lorentz} several textbooks and research
articles have appeared on the subject of RR. Many of them have criticized
aspects of the RR theory, and in particular the LAD and LL\ equations (for a
review see \cite{Rohrlich1965,Teitel1970a,Teitel1970b,parrot1987,parrot1993}%
, where one can find the discussion of the related problems). However,
despite contrary claims \cite%
{Spohn2000,Rohrlich2001,Rohrlich2000,Medina2006,Rohrlich2008,caldirola},
rigorous results are scarce \cite{Dorigo2008a}. In particular, most of
previous investigations concern the treatment of point charges. These are
usually based either on suitable asymptotic approximations or regularization
schemes to deal with intrinsic divergences of the point-charge model. On the
other hand, there is no obvious classical physical mechanism, consistent
with the SR-CE axioms, which can explain the appearance of a\emph{\ finite}
EM self-force acting on a point charge. This should arise as a consequence
of a \emph{finite delay time }occurring between the particle position at the
time of the generation of its EM self-field and the instantaneous particle
position. It is well-known, as discovered by Lorentz himself (Lorentz, 1892
\cite{Lorentz}; see also for example Landau and Lifschitz, 1951 \cite{LL})
that such a force can act on a charged particle only if the particle itself
is actually \emph{finite-size}. Therefore, although \textquotedblleft ad
hoc\textquotedblright\ models based on the adoption of a finite delay time
have been known for a long time (see for example the heuristic approach to
the RR problem by Caldirola, 1956 \cite{caldirola} leading to a delay-type
differential equation), the treatment of extended charge distributions
emerges as the only possible alternative, in analogy with the case of the
Debye screening problem in electrostatics \cite{Tessarotto2006}. In this
regard, a first approach in this direction is provided by the paper by
Nodvik (Nodvik,1964 \cite{Nodvik1964}), where a variational treatment for
point mass particles having finite-size charge distributions was developed.
However, charge and mass are expected to have the same support, as required,
for example, by the energy-momentum conservation law in both special and
general relativity. Therefore a fully consistent relativistic theory should
actually be formulated for\emph{\ finite-size particles}. From the analysis
of previous literature two important related problems arise:

\begin{itemize}
\item Issue \#1 - \emph{Existence of an exact variational RR equation: }this
refers to the lack of an exact RR equation, based on Hamilton variational
principle, even for classical point-particles (or point-masses). In fact,
previous approaches have all been based on approximate (i.e., asymptotic)
estimates. Example of this type leading to the well-known LAD equation
(Lorentz, Abraham and Dirac \cite{Lorentz,Yagh,Abraham1905,Dirac1938}) are
those due to Nodvik \cite{Nodvik1964} and Medina \cite{Medina2006}. A
critical aspect of the LAD equation, as well as of the related LL (Landau
and Lifschitz, 1951 \cite{LL}) equation, is that it does not satisfy a
variational principle in the customary sense, i.e., according to Axiom $\#2$
\cite{Dorigo2008a}. In particular, the resulting LAD equation is only
asymptotic and \emph{non-variational} in the sense of Axiom $\#2$. Instead,
the LL is non-variational, i.e., it \emph{does not} admit a variational
action at all. However, the problem arises whether, in the context of
special relativity, an \emph{exact RR equation} actually exists which holds
for suitable classical finite-size charged particles, and for Lorentzian
particles as a limiting case, namely finite-size charges having point-mass
distributions. Important related issues follow, such as the possibility for
the resulting equation to admit a \emph{standard Lagrangian form} in terms
of a \emph{non-local effective Lagrangian function}, and to be cast in an
equivalent \emph{conservative form, }as the divergence of an \emph{effective
stress-energy tensor.} Finally, the recovery of the customary LAD equation
in a suitable approximation must be verified.

\item Issue \#2 - \emph{Existence and uniqueness problem: }the second issue
is related to the consistency of the variational RR equation with the SR-CE
axioms and in particular with NDP. Therefore, the question arises whether an
existence and uniqueness theorem for the corresponding initial value problem
can be reached or not. Clearly the problem is relevant only for the exact RR
equation (yet to be established).
\end{itemize}

Clearly, the possible solution of these problems has potential wide-ranging
implications which are related to the description of relativistic dynamics
of systems of classical finite-size particles both in special and general
relativity.

\section{\label{sec:level2}Goals of the paper and scheme of the presentation}

The aim of the research program, of which the first part is reported here,
is to provide a consistent and exact theoretical formulation of the RR
problem for classical charged particles with \emph{finite-size charge and
mass distributions}, addressing precisely issues \#1 and \#2. In this paper
the case is considered of extended particles having mass and charge
distributions localized on the same support, identified with a surface shell
(see Section 3 for a complete rigorous definition). The result is obtained
without introducing any perturbative or asymptotic expansion for the
evaluation of EM self 4-potential and/or \textquotedblleft ad
hoc\textquotedblright\ regularization schemes for its point-particle limit.
In particular, finite-size charge distributions are introduced in order to
avoid intrinsic divergences (characteristic of the point-charge treatment)
and achieve an analytical description of the RR phenomena which is
consistent with the SR-CE axioms. A covariant representation for the EM self
4-potential is obtained, uniquely determined by the prescribed charge
current density. This allows us to point out the characteristic non-local
feature of the EM self-field, which is due to a causal retarded effect,
produced by the finite spatial extension of the charge. Here we shall
restrict the analysis to the treatment of charge and mass translational
motion, leaving the inclusion of rotational dynamics to a subsequent study.
Therefore, a suitable mathematical formulation of the problem is given, in
which rotational degrees of freedom are effectively excluded from the
present investigation. As a further result, it is proved that the exact RR
equation here obtained also holds for \emph{classical non-rotating
Lorentzian particles }(Lorentz, 1892\cite{Lorentz}), i.e., in the case in
which the mass is regarded as point-wise localized and only the charge has a
finite spatial extension. The approach here adopted is based on the
variational formulation for finite-size charged particles earlier pointed
out by Tessarotto \textit{et al.} \cite{Tessarotto2008c}, in turn relying on
the hybrid form of the synchronous variational principle \cite%
{Pozzo1998,Beklemishev1999}. A key feature of this variational principle is
the adoption of superabundant dynamic variables \cite{Cremaschini2006}(see
also related discussion in Sections 5 and 7). Due to the arbitrariness of
their definition, they can always be identified with the components of the
particle position and velocity 4-vectors $r^{\mu }$ and $u^{\mu }$. This
also implies that, by construction, the variational functional necessarily
satisfies the property of covariance and MLC. Then, the corresponding E-L
equations yield both the RR equation and also the required physical
realizability constraints for $r^{\mu }$ and $u^{\mu }$, which allow one to
identify them with physical observables.

The paper is organized as follows. In Section 3 we present the derivation of
the charge and mass current densities for the particle model adopted, while
in Section 4 the exact solution for the EM self 4-potential generated by the
non-rotating charge distribution is constructed (Lemma 1). On the basis of
this result an explicit integral representation is obtained for the EM
self-potential (Lemma 2). Subsequently, in Section 5 we proceed in detail to
the construction of the variational functional. In particular, by making use
of Lemma 3, the contributions from the EM-coupling with both the EM
self-field (Subsection 5.1) and the external EM field (Subsection 5.2), as
well as the inertial mass contribution (Subsection 5.4) are determined.
Then, in Section 6 the resulting variational Lagrangian is derived. In
Section 7 the variational formulation for finite-size particles is
presented, based on a synchronous variational principle (THM.1). As a
fundamental consequence, it is found that the RR equation is a covariant
second-order \emph{delay differential equation} which fulfills all SR-CE
axioms and in particular ECP, GIP and MLC. The equation is proved to apply
also to the particular case of Lorentzian particles. Then, in Section 8 the
RR equation is shown to admit both standard Lagrangian and conservative
forms (THM.2). Section 9 deals instead with the asymptotic behavior of the
RR equation, showing that in the short delay-time approximation it recovers
the customary LAD equation, while not admitting the point-charge limit
(THM.3). Finally, in Section 10 it is proved that, under suitable physical
assumptions, the RR equation here obtained fulfills also NDP and,
consequently, admits a well-posed initial value problem (i.e., there is an
\emph{existence and uniqueness theorem}; see THM.4).

\section{Charge and mass current densities}

In this section we define the particle model, prescribing its mass and
charge distributions, and determine the corresponding covariant expressions
for the charge and mass current densities, both needed for the subsequent
developments. Here we consider the treatment in the special relativity
setting.

By definition, the particle is characterized by a positive constant rest
mass $m_{o}$ and a non-vanishing constant charge $q$, with surface mass and
charge densities $\rho _{m}$ and $\rho _{c}$ respectively. We shall assume
that the mass and charge distributions have supports $\partial \Omega _{m}$
and $\partial \Omega _{\sigma }$. To define the particle mass and charge
distributions on $\partial \Omega _{m}$ and $\partial \Omega _{\sigma },$
let us assume initially that in a time interval $[-\infty ,t_{o}]$ the
particle is at rest with respect to an inertial frame (i.e., that external
forces acting on the particle vanish identically). As a consequence, by
assumption in the subset of the space-time $\mathcal{M}^{4}\subseteq \mathbb{%
R}^{4}$ in which $t\in \lbrack -\infty ,t_{o}],$ there is an inertial frame
in which both the particle mass and charge distributions are at rest
(particle rest-frame $\mathcal{R}_{o}$). In this frame, we shall assume that
there exists a point, hereafter referred to as \textit{center of symmetry
(COS)}, whose position 4-vector $r_{COS}^{\mu }\equiv (ct,\mathbf{r}_{o})$
spans the Minkowski space-time $\mathcal{M}^{4}\subseteq \mathbb{R}^{4}$ and
with respect to which:

1) $\partial \Omega _{\sigma }$ and $\partial \Omega _{m}$ are stationary
spherical surfaces of radii $\sigma >0$ and $\sigma _{m}>0$ of equations $%
\left( \mathbf{r-r}_{o}\right) ^{2}=\sigma ^{2}$ and $\left( \mathbf{r-r}%
_{o}\right) ^{2}=\sigma _{m}^{2};$

2) the particle is \emph{quasi-rigid}, i.e., the mass and charge
distributions are stationary and spherically-symmetric respectively on $%
\partial \Omega _{m}$ and $\partial \Omega _{\sigma }$\footnote{%
In order to warrant the condition of rigidity in a manner consistent with
the SR-CE Axioms, following the literature a possibility is to assume that
the extended particle is acted upon by a local non-EM force
\textquotedblleft whose precise nature is left
unspecified\textquotedblright\ (see Nodvik \cite{Nodvik1964} and further
references indicated there).}$;$

3) in addition, consistent with the principle of energy-momentum
conservation (see further discussion below), we shall assume the
distributions of mass and charge densities to have the same support \emph{\ }%
$\partial \Omega _{\sigma }\equiv \partial \Omega _{m}$, hence letting
\begin{equation}
\sigma _{m}=\sigma .  \label{SUPPORT}
\end{equation}
Finally, the case in which the\ mass is considered localized point-wise (%
\emph{Lorentzian particle}) is recovered letting $\sigma _{m}\neq \sigma,$
with $\sigma >0$ and $\sigma _{m}=0$. \ In both cases the particle mass and
charge distributions remain uniquely defined in any reference frame for
arbitrary particle motion.

In this paper, we are concerned only with the investigation of the EM RR
phenomenon on the translational dynamical motion of the charged particle.
Hence, we require that the mass density (and, as a consequence, also the
charge density) does not possess pure spatial rotation, nevertheless still
allowing for space-time rotations (i.e., Thomas precession, see below). For
definiteness, let us introduce here the Euler angles $\alpha (s)\equiv
\left\{ \varphi (s),\vartheta (s),\psi (s)\right\} $ which define the
orientation of the body-axis system $K^{\prime }$ with respect to the rest
system $K$ (according to the notations used by Nodvik \cite{Nodvik1964}).
Introducing the generalized velocities $\frac{d\alpha \left( s\right) }{ds}%
\equiv \left\{ \frac{d\varphi }{ds},\frac{d\vartheta }{ds},\frac{d\psi }{ds}%
\right\} ,$ the condition of \emph{vanishing mass and charge spatial rotation%
} in a time interval $I\subseteq \mathbb{R}$ is thus prescribed imposing
that the particular solution
\begin{eqnarray}
\alpha (s) &=&\alpha _{o},  \notag \\
\frac{d\alpha \left( s\right) }{ds} &\equiv &0,  \label{nonnrot}
\end{eqnarray}%
holds for all $s\in I.$ For a physical motivation for this assumption we
refer to the discussion reported by Yaghjian \cite{Yagh}.

\bigskip

Having specified the physical properties of the particle by means of the
mass and charge distributions, we can now move on to obtaining the covariant
expression for the corresponding charge and mass current densities. Since
the charge and the mass have the same support, the mathematical derivation
is formally the same for both of them. For convenience we start with the
charge current $j^{\mu }(r)$, introducing for it the representation used by
Nodvik. For definiteness, let us denote \cite{Nodvik1964}
\begin{eqnarray*}
s &\equiv &\emph{\ proper time of the COS,} \\
r^{\mu }(s) &\equiv &\emph{COS 4-position,} \\
\zeta ^{\mu } &\equiv &\emph{\ charge element 4-position.}
\end{eqnarray*}%
Then, we define the displacement vector $\xi ^{\mu }$ as follows:%
\begin{equation}
\xi ^{\mu }\equiv \varsigma ^{\mu }-r^{\mu }(s),
\end{equation}%
from which we also have that $\varsigma ^{\mu }=r^{\mu }(s)+\xi ^{\mu }.$
The physical meaning of the 4-vector $\xi ^{\mu }$ is that of a displacement
between the particle\textbf{\ }COS and its boundary, where the charge is
located. According to this representation, $\xi ^{\mu }$ is subject to the
following two constraints \cite{Nodvik1964}:%
\begin{eqnarray}
\xi ^{\mu }\xi _{\mu } &=&-\sigma ^{2},  \label{eee1} \\
\xi _{\mu }u^{\mu }(s) &=&0,  \label{eee2}
\end{eqnarray}%
where%
\begin{equation}
u^{\mu }(s)\equiv \frac{d}{ds}r^{\mu }(s)
\end{equation}%
is the 4-velocity of the COS. The first equality (\ref{eee1}) defines the
boundary $\partial \Omega _{\sigma }=\partial \Omega _{m}$. The second
constraint (\ref{eee2}) represents instead the constraint of rigidity for
the particle. This implies that in the particle rest frame the 4-vector $\xi
^{\mu }$ has only spatial components. We can use the information from Eq.(%
\ref{eee1}) to define the internal and the external domains with respect to
the mass and charge distributions. In particular, if we define a generic
displacement 4-vector $X^{\mu }\in M^{4}$ as
\begin{equation}
X^{\mu }=r^{\mu }-r^{\mu }\left( s\right),  \label{aaa}
\end{equation}%
which is subject to the constraint%
\begin{equation}
X^{\mu }u_{\mu }(s)=0,  \label{bbb}
\end{equation}%
then the following relations hold:%
\begin{eqnarray}
X^{\mu }X_{\mu } &\leq &-\sigma ^{2}\emph{\ : external domain,}
\label{extdom} \\
X^{\mu }X_{\mu } &>&-\sigma ^{2}\emph{\ : internal domain,}  \notag \\
X^{\mu }X_{\mu } &=&\xi ^{\mu }\xi _{\mu }=-\sigma ^{2}\emph{\ : boundary
location.}  \notag
\end{eqnarray}%
To derive the current density 4-vector corresponding to the spherical
charged shell we follow the presentation by Nodvik \cite{Nodvik1964}.
Consider first the charge-current density $\Delta j^{\mu }(r)$ corresponding
to a charge element $\Delta q$ on the shell. This is expressed as follows:%
\begin{equation}
\Delta j^{\mu }(r)=c\Delta q\int_{1}^{2}d\zeta ^{\mu }\delta ^{4}\left(
r^{\mu }-\zeta ^{\mu }\right) =c\Delta q\int_{-\infty }^{+\infty }ds\left[
u^{\mu }+\frac{d\xi ^{\mu }}{ds}\right] \delta ^{4}\left( x^{\mu }-\xi ^{\mu
}\right) ,
\end{equation}%
where%
\begin{equation}
x^{\mu }=r^{\mu }-r^{\mu }\left( s\right) .  \label{xx}
\end{equation}%
Note that, for the simplicity of the notation, here and in the rest of the
paper the symbol $r$ stands for the generic 4-vector $r^{\alpha }$ when used
as an argument of a function. Since the charge does not possess any pure
spatial rotation, the relation%
\begin{equation}
\frac{d\xi ^{\mu }}{ds}=\Gamma u^{\mu }
\end{equation}%
holds, where $\Gamma \equiv -\left( \frac{du_{\alpha }}{ds}\xi ^{\alpha
}\right) $ carries the effect associated with the Thomas precession\cite%
{Nodvik1964}. The expression for $\Delta j^{\mu }(r)$ then becomes%
\begin{equation}
\Delta j^{\mu }(r)=c\Delta q\int_{-\infty }^{+\infty }dsu^{\mu }\left[
1+\Gamma \right] \delta ^{4}\left( x^{\mu }-\xi ^{\mu }\right) .
\end{equation}%
To compute the total current of the charged shell we express the charge
element $\Delta q$ according to the constraint (\ref{eee2}) as follows: %
$\Delta q=qf(\left\vert \xi \right\vert )\delta (\xi ^{\alpha }u_{\alpha
}(s))d^{4}\xi ,$ 
where $d^{4}\xi $ is the 4-volume element in the $\xi $-space. Moreover, $%
f(\left\vert \xi \right\vert )$ is referred to as the form factor, which
describes the charge distribution of the moving body. In particular, for a
spherically symmetric distribution this has the following representation:%
\begin{equation}
f(\left\vert \xi \right\vert )=\frac{1}{4\pi \sigma ^{2}}\delta (\left\vert
\xi \right\vert -\sigma ),
\end{equation}%
where $\left\vert \xi \right\vert \equiv \left\vert \sqrt{\xi ^{\mu }\xi
_{\mu }}\right\vert .$ The total current density $j^{\mu }(r)$ can therefore
be obtained by integrating $\Delta j^{\mu }(r)$ over $d^{4}\xi $. We get%
\begin{eqnarray}
j^{\mu }(r) &\equiv &qc\int_{-\infty }^{+\infty }dsu^{\mu
}\int_{1}^{2}d^{4}\xi f(\left\vert \xi \right\vert )\delta (\xi ^{\alpha
}u_{\alpha })\left[ 1+\Gamma \right] \delta ^{4}\left( x^{\mu }-\xi ^{\mu
}\right) =  \notag \\
&=&qc\int_{-\infty }^{+\infty }dsu^{\mu }f(\left\vert x\right\vert )\delta
(x^{\alpha }u_{\alpha })\left[ 1+\Gamma \right] ,
\end{eqnarray}%
where%
\begin{equation}
f(\left\vert x\right\vert )=\frac{1}{4\pi \sigma ^{2}}\delta (\left\vert
x\right\vert -\sigma )
\end{equation}%
with $\left\vert x\right\vert \equiv \left\vert \sqrt{x^{\mu }x_{\mu }}%
\right\vert .$ Then we notice that%
\begin{equation}
\delta (x^{\alpha }u_{\alpha }(s))=\frac{1}{\left\vert \frac{d\left[
x^{\alpha }u_{\alpha }\right] }{ds}\right\vert }\delta (s-s_{1})=\frac{1}{%
\left\vert 1+\Gamma \right\vert }\delta (s-s_{1}),
\end{equation}%
where by definition $s_{1}$ is the root of the algebraic equation%
\begin{equation}
u_{\mu }(s_{1})\left[ r^{\mu }-r^{\mu }\left( s_{1}\right) \right] =0.
\label{s1}
\end{equation}%
Combining these relations, it follows that the integral covariant expression
for the charge current density is given by%
\begin{equation}
j^{\mu }(r)=\frac{qc}{4\pi \sigma ^{2}}\int_{-\infty }^{+\infty }dsu^{\mu
}(s)\delta (\left\vert x\right\vert -\sigma )\delta (s-s_{1}).  \label{intj}
\end{equation}

Finally, an analogous expression for the mass current density $j_{mass}^{\mu
}(r)$ can be easily obtained from $j^{\mu }(r)$ by replacing the total
charge $q$ with the total mass $m_{o}$, thus giving%
\begin{equation}
j_{mass}^{\mu }(r)=\frac{m_{o}c}{4\pi \sigma ^{2}}\int_{-\infty }^{+\infty
}dsu^{\mu }(s)\delta (\left\vert x\right\vert -\sigma )\delta (s-s_{1}).
\label{mass_current}
\end{equation}%
We remark that in both equations (\ref{intj}) and (\ref{mass_current}):

1) the dependence in terms of the 4-position $r$ enters explicitly through $%
|x|=|r^{\mu }-r^{\mu }\left( s\right) |$ in the form factor and implicitly
through the root $s_{1}$;

2) consistent with assumption (\ref{nonnrot}), possible charge and mass
spatial rotations have been set to be identically zero.

\bigskip

\section{EM self 4-potential - Case of non-rotating charge distribution}

A prerequisite for the subsequent developments is the determination of the
EM self-potential ($A_{\mu }^{(self)}$) produced by the spherical charged
particle shell here introduced. In principle the problem could be formally
treated by solving the Maxwell equations with the 4-potential written in
terms of a suitable Green function according to standard methods.
Remarkably, the solution can also be achieved in a more straightforward way
based on the relativity principle and the covariance of Maxwell's equations.
This implies the possibility of obtaining a covariant representation of the
EM 4-vector in a generic reference system once its definition is known in a
particular reference frame. The approach is analogous to the derivation
presented by Landau and Lifschitz \cite{LL} for the treatment of a point
charge. The solution is provided by the following Lemmas.

\bigskip

\textbf{Lemma 1 - Covariant representation for }$A_{\mu }^{(self)}(r)$

\emph{Given validity of the assumptions on the particle structure introduced
in the previous section and the results obtained for the current density,
the following statements hold:}

L1$_{1}:$ \emph{Particle at rest in an inertial frame.}

\emph{Let us assume that the particle is at rest in an inertial frame }$%
S_{0} $ \emph{and, according to (\ref{nonnrot}), is non-rotating in this
frame.} \emph{By definition, in} $S_{0}$ \emph{the 4-vector potential of the
self-field is written as\ }$A_{\mu }^{(self)}(r)=A_{S_{0}\mu
}^{(self)}(r)\equiv \left\{ \Phi ^{(self)},\mathbf{0}\right\} ,$\emph{\
where }%
\begin{equation}
\Phi ^{(self)}(\mathbf{r},t)=\left\{
\begin{array}{ccc}
\frac{q}{R} &  & (R\geq \sigma ), \\
\frac{q}{\sigma } &  & (R<\sigma ),%
\end{array}%
\right.  \label{REST-FTAME REPRESENTATION}
\end{equation}%
\emph{(rest-frame representation) denote respectively the external and
internal solutions with respect to the boundary of the shell. Here }%
\begin{eqnarray}
R &\equiv &\left\vert \mathbf{R}\right\vert , \\
\mathbf{R} &=&\mathbf{r}-\mathbf{r}\left( t^{\prime }\right) ,
\end{eqnarray}%
\emph{with }$r^{\mu }=(ct,\mathbf{r}),$\emph{\ }$r^{\prime \mu }=(ct^{\prime
},\mathbf{r}^{\prime }\equiv \mathbf{r}\left( t^{\prime }\right) ),$ \emph{%
and} $\mathbf{r},\mathbf{r}^{\prime }\equiv \mathbf{r}\left( t^{\prime
}\right) $ \emph{being respectively} \emph{a generic position }$3$\emph{%
-vector of }$%
\mathbb{R}
^{3}$\emph{\ and the (stationary) position 3-vector of the particle COS. It
follows that }$\Phi ^{(self)}(\mathbf{r},t)$ \emph{can be equivalently
represented as }%
\begin{equation}
\Phi ^{(self)}(\mathbf{r},t)=\left\{
\begin{array}{ccc}
\frac{q}{c(t-t^{\prime })}\equiv \frac{q}{R} &  & (R\geq \sigma ), \\
\frac{q}{c(t-t^{\prime })}\equiv \frac{q}{\sigma } &  & (R<\sigma ),%
\end{array}%
\right.  \label{REST-FTAME REPRESENTATION-1}
\end{equation}%
\emph{where }$t_{ret}\equiv t-t^{\prime }$ \emph{is} \emph{the following
positive root}%
\begin{equation}
t_{ret}\equiv t-t^{\prime }=\left\{
\begin{array}{ccc}
t_{ret}^{(ext)}\equiv \pm \frac{R}{c} &  & (R\geq \sigma ), \\
t_{ret}^{(int)}\equiv \pm \frac{\sigma }{c} &  & (R<\sigma ).%
\end{array}%
\right.  \label{DELAY-TIME}
\end{equation}

L1$_{2}:$ \emph{Particle with inertial motion in an arbitrary inertial frame.%
}

\emph{Let us assume that when the particle is referred to an arbitrary
inertial frame }$S_{I}$\emph{\ it has a constant }$4-$\emph{velocity} $%
u^{\alpha }\equiv $\emph{\ }$\frac{dr^{\mu }(s^{\prime })}{ds^{\prime }}.$
\emph{Then, let us require that} $t_{ret}\equiv t-t^{\prime }$ \emph{is}
\emph{the positive root of the delay-time equation}%
\begin{equation}
\emph{\ }\widehat{R}^{\alpha }\emph{\ }\widehat{R}_{\alpha }=\rho ^{2},
\label{DELAY TIME EQUATION-FINAL}
\end{equation}%
\emph{with }$\widehat{R}^{\alpha }$ \emph{being the bi-vector}%
\begin{equation}
\emph{\ }\widehat{R}^{\alpha }=r^{\alpha }-r^{\alpha }(t^{\prime })
\label{bivector}
\end{equation}%
\emph{and}%
\begin{equation}
\rho ^{2}=\left\{
\begin{array}{ccc}
0 &  & (X^{\alpha }X_{\alpha }\leq -\sigma ^{2}), \\
\rho ^{2}\equiv \sigma ^{2}\left[ 1+\frac{X^{\alpha }X_{\alpha }}{\sigma ^{2}%
}\right] &  & (X^{\alpha }X_{\alpha }>-\sigma ^{2}),%
\end{array}%
\right.  \label{EQUATION FOR RO -FINAL}
\end{equation}%
\emph{where the displacement vector }$X^{\alpha }$ \emph{is defined by Eqs.(%
\ref{aaa}) and (\ref{bbb}). For consistency, Eq.(\ref{EQUATION FOR RO -FINAL}%
) provides the solution Eq.(\ref{DELAY-TIME}) when evaluated in the COS
comoving frame.}

\emph{It follows that in the reference frame }$S_{I}$ \emph{the EM self
4-potential have the internal and external solutions}
\begin{equation}
A_{\mu }^{(self)}(r)=\left\{
\begin{array}{ccc}
\left. q\frac{u_{\mu }}{\widehat{R}^{\alpha }u_{\alpha }}\right\vert
_{t_{ret}=t_{ret}^{(ext)}} &  & (X^{\alpha }X_{\alpha }\leq -\sigma ^{2}),
\\
\left. q\frac{u_{\mu }}{\widehat{R}^{\alpha }u_{\alpha }}\right\vert
_{t_{ret}=t_{ret}^{(int)}} &  & (X^{\alpha }X_{\alpha }>-\sigma ^{2}),%
\end{array}%
\right.  \label{aaaa SOLUTION-1}
\end{equation}%
\emph{where }$\widehat{R}^{\alpha }$\emph{\ is given by Eq.(\ref{bivector})}$%
.$

L1$_{3}:$ \emph{Particle with a non-inertial motion in an arbitrary frame.}

\emph{Let us assume that the same particle is now referred to an arbitrary
frame in which it has a time-dependent velocity }$u_{\mu }(t^{\prime })$%
\emph{. In this frame the EM self 4-potential }$A_{\mu }^{(self)}(r)$\emph{\
takes the form:}%
\begin{equation}
A_{\mu }^{(self)}(r)=\left\{
\begin{array}{ccc}
\left. q\frac{u_{\mu }(t^{\prime })}{\widehat{R}^{\alpha }u_{\mu }(t^{\prime
})}\right\vert _{t_{ret}=t_{ret}^{(ext)}} &  & (X^{\alpha }X_{\alpha }\leq
-\sigma ^{2}), \\
\left. q\frac{u_{\mu }(t^{\prime })}{\widehat{R}^{\alpha }u_{\mu }(t^{\prime
})}\right\vert _{t_{ret}=t_{ret}^{(int)}} &  & (X^{\alpha }X_{\alpha
}>-\sigma ^{2}),%
\end{array}%
\right.  \label{DUFFERENTIAL REPRESENTATION-2}
\end{equation}%
\emph{where }$u_{\mu }(t^{\prime })$\emph{\ is the 4-velociy of the COS with
4-position }$r^{\alpha }(t^{\prime }),$\emph{\ i.e.,}
\begin{equation}
u_{\mu }(t^{\prime })\equiv \frac{dr^{\beta }(t^{\prime })}{ds^{\prime }}%
=\gamma (t^{\prime })\frac{dr^{\beta }(t^{\prime })}{cdt^{\prime }},
\end{equation}%
\emph{and} $t_{ret}^{(ext)},$ $t_{ret}^{(int)}$ \emph{are the positive roots
of the delay-time equation (\ref{DELAY TIME EQUATION-FINAL}).}

\emph{Proof} - \ \ L1$_{1})$\emph{\ }If the particle is at rest in an
inertial frame $S_{0}$, from the form of the charge density (\ref{intj}) and
the condition of non-rotation (\ref{nonnrot}), the EM self\ 4-potential is
stationary in $S_{0}$. Hence it takes necessarily the form $A_{\mu
}^{(self)}(r)=A_{S_{0}\mu }^{(self)}(r)\equiv \left\{ \Phi ^{(self)},\mathbf{%
0}\right\} .$ Thus, denoting
\begin{eqnarray}
R &\equiv &\left\vert \mathbf{R}\right\vert , \\
\mathbf{R} &=&\mathbf{r}-\mathbf{r}\left( t-\frac{\left\vert \mathbf{r}-%
\mathbf{r}(t-\frac{R}{c})\right\vert }{c}\right) ,
\end{eqnarray}%
with $\mathbf{r}$ a generic position $3$-vector of $%
\mathbb{R}
^{3}$ and $\mathbf{r}(t^{\prime })\equiv \mathbf{r}(t-\frac{R}{c})$ the
retarded-time position 3-vector, \ $\Phi ^{(self)}$ is written as%
\begin{equation}
\Phi ^{(self)}(\mathbf{r},t)=\left\{
\begin{array}{ccc}
\frac{q}{R} &  & (R\geq \sigma ), \\
\frac{q}{\sigma } &  & (R<\sigma ).%
\end{array}%
\right.
\end{equation}%
In other words, in the external/internal sub-domains (respectively defined
by the inequalities $R\geq \sigma $ and $R<\sigma $) the ES potential $\Phi
^{(self)}$ coincides with the ES potential of a point charge and a constant
potential. In terms of the delay time $t_{ret}=t=t^{\prime }$ determined by
Eq.(\ref{DELAY-TIME}) it is immediate to prove Eq.(\ref{REST-FTAME
REPRESENTATION-1}).

L1$_{2})$ Next, let us consider the same particle referred to an arbitrary
inertial frame $S_{I}$ in which the COS position vector $r^{\alpha
}(s^{\prime })$ has a constant velocity
\begin{equation}
u_{\alpha }\equiv u^{\alpha }(s^{\prime })=\frac{d}{ds^{\prime }}r^{\alpha
}(s^{\prime })=const.  \label{4-velocity}
\end{equation}%
Since by definition $A_{\mu }^{(self)}(r)$ is a covariant 4-vector, its form
in $S_{I}$ is simply obtained by applying a Lorentz transformation \cite{LL}
according to Eq.(\ref{4-velocity}). This requires
\begin{equation}
A_{\mu }^{(self)}(r)=q\frac{u_{\mu }}{\widehat{R}^{\alpha }u_{\alpha }},
\end{equation}%
where $\widehat{R}^{\alpha }=r^{\alpha }-r^{\alpha }(s^{\prime })$. Denoting
$s^{\prime }\equiv s^{\prime }(t^{\prime })$ and $r^{\alpha }(s^{\prime
})\equiv (ct^{\prime },\mathbf{r}(t^{\prime })),$ let us now impose that $%
t-t^{\prime }$ is the positive root of the delay-time equation (\ref{DELAY
TIME EQUATION-FINAL}). The external and internal solutions in this case are
given respectively by Eq.(\ref{aaaa SOLUTION-1}), as can be seen by noting
that when $u_{\mu }=(1,\mathbf{0})$ the correct external and internal
solutions (\ref{REST-FTAME REPRESENTATION}) are recovered.

L1$_{3})$\emph{\ }The proof of the third statement is a basic consequence of
the principle of relativity and of the covariance of the Maxwell equations.
In fact we notice that both the solution (\ref{aaaa SOLUTION-1}) for the
4-vector potential and Eq.(\ref{DELAY TIME EQUATION-FINAL}) for the delay
time, which have been obtained for the specific case of an inertial frame,
are already written in covariant form by means of the 4-vector notation.
Hence, according to the principle of relativity, this solution is valid in
any reference system related by a Lorentz transformation, and for a generic
form of the 4-velocity $u_{\mu }$ (cf Landau and Lifshitz \cite{LL}).

\textbf{Q.E.D.}

\bigskip

We remark that Eq.(\ref{DUFFERENTIAL REPRESENTATION-2}) provides an exact
representation (defined up to a gauge transformation) for the EM self
4-potential generated by the non-rotating finite-size charge considered here.

On the base of the conclusions of Lemma 1 it follows that $A_{\mu
}^{(self)}(r)$ can also be represented by means of an equivalent integral
representation as proved by the following Lemma.\newline

\textbf{Lemma 2 - Integral representation for }$A_{\mu }^{(self)}(r)$

\emph{Given validity of Lemma 1, the EM self 4-potential Eq.(\ref%
{DUFFERENTIAL REPRESENTATION-2}) admits the equivalent integral
representation }%
\begin{equation}
A_{\mu }^{(self)}(r)=2q\int_{1}^{2}dr_{\mu }^{\prime }\delta (\widehat{R}%
^{\alpha }\widehat{R}_{\alpha }-\rho ^{2}),
\label{AAAAA=INTEGRAL
REPRESENTATION}
\end{equation}%
\emph{with }$\rho ^{2}$\emph{\ defined by Eq.(\ref{EQUATION FOR RO -FINAL})
and }$r_{\mu }^{\prime }\equiv r_{\mu }\left( s^{\prime }\right) $\emph{.}

\emph{Proof} - In fact in the external and internal domains
\begin{equation}
\delta (\widehat{R}^{\alpha }\widehat{R}_{\alpha }-\rho ^{2})=\left\{
\begin{array}{ccc}
\frac{\delta (s-s^{\prime })}{2\left\vert \widehat{R}_{\alpha }\frac{%
dr^{\prime \alpha }}{ds^{\prime }}\right\vert } &  & (X^{\alpha }X_{\alpha
}\leq -\sigma ^{2}), \\
\frac{\delta (s-s^{\prime })}{\left\vert 2\widehat{R}_{\alpha }\frac{%
dr^{\prime \alpha }}{ds^{\prime }}+\frac{d\rho ^{2}}{ds}\right\vert } &  &
(X^{\alpha }X_{\alpha }>-\sigma ^{2}),%
\end{array}%
\right.
\end{equation}%
where $\frac{d\rho ^{2}}{ds}=\frac{dX^{\alpha }X_{\alpha }}{ds^{\prime }}%
=2X_{\alpha }u_{\alpha }(s^{\prime })\equiv 0$ because of Eq.(\ref{bbb}),
while $s^{\prime }$ is determined by the delay-time equation (\ref{DELAY
TIME EQUATION-FINAL})$.$ Hence, Eq.(\ref{AAAAA=INTEGRAL REPRESENTATION})
manifestly implies Eq.(\ref{DUFFERENTIAL REPRESENTATION-2}).

\textbf{Q.E.D.}

\bigskip

\section{The action integral}

In this section we derive the Hamilton action functional suitable for the
variational treatment of finite-size charged particles introduced here and
the investigation of their dynamics. As indicated in Section 3, the
contributions due to pure spatial charge and mass rotations will be ignored.
In this case, the action integral is conveniently expressed in hybrid
superabundant variables (see Tessarotto \textit{et al.} \cite%
{Cremaschini2006}) as follows:%
\begin{equation}
S_{1}(r,u,\chi ,\left[ r\right] )=S_{M}(r,u)+S_{C}^{\left( self\right) }(r,%
\left[ r\right] )+S_{C}^{\left( ext\right) }(r)+S_{\chi }(u,\chi ),
\label{pointp}
\end{equation}%
where $S_{M}$, $S_{C}^{\left( self\right) }$, $S_{C}^{\left( ext\right) }$
and $S_{\chi }$ are respectively the inertial mass, the EM-coupling with the
self and external fields, and the kinematic constraint contributions. For
what concerns the notation, here $r$ and $u$ represent \textit{local}
depepndencies with respect to the 4-vector position $r^{\mu }$ and the
4-velocity $u^{\mu }$, $\left[ r\right] $ stands for \textit{non-local}
dependencies on the 4-vector position $r^{\mu }$, while $\chi \equiv \chi
(s) $ is a Lagrange multiplier (see also below and the related discussion in
THM.1 of Section 7).

Before addressing the explicit evaluation of $S_{1}(r,u,\chi ,\left[ r\right]
)$ we prove the following preliminary Lemma concerning the transformation
properties of 4-volume elements under Lorentz transformations.

\bigskip

\textbf{Lemma 3 - Lorentz transformations and 4-volume elements}

\emph{Let us consider a Lorentz transformation (Lorentz boost) from an
inertial reference frame }$S_{I}$ \emph{to a reference frame }$S_{NI}$ \emph{%
whose origin has 4-velocity }$u_{\mu }(s_{2})$\emph{\ with respect to }$%
S_{I} $, \emph{with }$s_{2}$\emph{\ being considered here an arbitrary
proper time independent of $r^{\mu }\in S_{I}$. By assumption }$u_{\mu
}(s_{2})$ \emph{is constant both with respect to the 4-positions }$r^{\mu
}\in S_{I}$\emph{\ and }$r^{\prime \mu }\in S_{NI}$\emph{\ in the two
reference frames. The relationship between the two 4-vectors }$r^{\mu }\in
S_{I}$\emph{\ and }$r^{\prime \mu }\in S_{NI}$\emph{\ is expressed by the
transformation law\cite{Jackson}}%
\begin{equation}
r^{\prime \mu }=\Lambda _{\nu }^{\mu }\left( u_{\mu }(s_{2})\right) r^{\nu },
\label{Lboost}
\end{equation}%
\emph{where }$\Lambda _{\nu }^{\mu }\left( u_{\mu }(s_{2})\right) $\emph{\
is the matrix of the Lorentz boost, which by definition depends only on the
relative 4-velocity }$u_{\mu }(s_{2})$\emph{\ between }$S_{I}$\emph{\ and }$%
S_{NI}$\emph{. Then it follows that the 4-volume element }$d\Omega \in S_{I}$%
\emph{\ is invariant with respect to the Lorentz boost (\ref{Lboost}), in
the sense: }%
\begin{equation}
d\Omega =d\Omega ^{\prime },  \label{dO}
\end{equation}%
\emph{with }$d\Omega ^{\prime }\in S_{NI}$ \emph{denoting the corresponding
volume element in the transformed frame }$S_{NI}$\emph{.}

\emph{Proof} - The proof of this statement follows by considering the
general transformation property of volume elements under arbitrary change of
coordinates. Consider the invariant 4-volume element $d\Omega \in S_{I}$ and
assume a Minkowski metric tensor. By definition \cite{LL}, for a generic
change of reference frame the volume element transforms according to the law%
\begin{equation}
d\Omega =\frac{1}{J}d\Omega ^{\prime },  \label{voltran}
\end{equation}%
where%
\begin{equation}
J\equiv \left\vert \frac{\partial r^{\mu }}{\partial r^{\prime \mu }}%
\right\vert
\end{equation}%
is the Jacobian of the corresponding coordinate transformation. In the case
considered here, the Lorentz boost (\ref{Lboost}) is described by the matrix
$\Lambda _{\nu }^{\mu }\left( u_{\mu }(s_{2})\right) $\emph{\ }which depends
only on the 4-velocity $u_{\mu }(s_{2}),$ by assumption independent of the
coordinates $r^{\nu }$ and $r^{\prime \nu }$. It follows that $J\equiv 1$,
implying in turn Eq.(\ref{dO}).

\textbf{Q.E.D.}

\bigskip

We can now proceed to evaluate the various contributions to the action
integral $S_{1}(r,u,\chi ,\left[ r\right] )$ defined in Eq.(\ref{pointp}).

$\bigskip $

\subsection{$S_{C}^{\left( self\right) }(r,\left[ r\right] )$: EM coupling
with the self-field}

The action integral $S_{C}^{\left( self\right) }(r,\left[ r\right] )$
containing the coupling between the EM self-field and the electric 4-current
is of critical importance. For this reason and for the sake of clarity in
this subsection the steps of its evaluation are reported in detail.
According to the standard approach \cite{LL}, $S_{C}^{\left( self\right) }$
is defined as the 4-scalar%
\begin{equation}
S_{C}^{(self)}(r,\left[ r\right] )=\int_{1}^{2}d\Omega \frac{1}{c^{2}}%
A^{(self)\mu }(r)j_{\mu }\left( r\right) ,  \label{act1}
\end{equation}%
where $A^{(self)\mu }(r)$ is given by Eq.(\ref{AAAAA=INTEGRAL REPRESENTATION}%
), $j_{\mu }\left( r\right) $ by Eq.(\ref{intj}) and $d\Omega $ is the
invariant 4-volume element. In particular, in an inertial frame $S_{I}$ with
Minkowski metric tensor $\eta _{\mu \nu }$, this can be represented as%
\begin{equation}
d\Omega =cdtdxdydz,
\end{equation}%
where $\left( x,y,z\right) $ are orthogonal Cartesian coordinates. The
functional can be equivalently represented as%
\begin{eqnarray}
S_{C}^{(self)}(r,\left[ r\right] ) &=&\frac{q}{4\pi \sigma ^{2}c}%
\int_{1}^{2}d\Omega A^{(self)\mu }(r)\int_{-\infty }^{+\infty }ds_{2}\delta
(s_{2}-s_{1})\times  \notag \\
&&\times \int_{-\infty }^{+\infty }dsu^{\mu }(s)\delta (\left\vert x\left(
s\right) \right\vert -\sigma )\delta (s-s_{2}),  \label{long2}
\end{eqnarray}%
where $s_{1}$ is the root of the equation%
\begin{equation}
u_{\mu }(s_{1})\left[ r^{\mu }-r^{\mu }\left( s_{1}\right) \right] =0.
\end{equation}%
Because of the principle of relativity, the integral (\ref{act1}) can be
evaluated in an arbitrary reference frame. The explicit calculation of the
integral (\ref{act1}) is then achieved, thanks to Lemma 3, by invoking the
Lorentz boost (\ref{Lboost}) to the reference frame $S_{NI}$ moving with
4-velocity $u_{\mu }(s_{2})$. In this frame, by construction $d\Omega
^{\prime }=cdt^{\prime }dx^{\prime }dy^{\prime }dz^{\prime }\equiv d\Omega $%
. In particular, introducing the spherical spatial coordinates $\left(
ct^{\prime },\rho ^{\prime },\vartheta ^{\prime },\varphi ^{\prime }\right) $
it follows that the transformed spatial volume element can also be written
as $cdt^{\prime }dx^{\prime }dy^{\prime }dz^{\prime }\equiv cdt^{\prime
}d\rho ^{\prime }d\vartheta ^{\prime }d\varphi ^{\prime }\rho ^{\prime
2}\sin \vartheta ^{\prime }.$ In this frame the previous scalar equation
becomes%
\begin{equation}
u_{\mu }^{\prime }(s_{1})\left[ r^{\prime \mu }-r^{\prime \mu }\left(
s_{1}\right) \right] =0.  \label{bis}
\end{equation}%
On the other hand, performing the integration with respect to $s_{2}$ in Eq.(%
\ref{long2}), it follows that necessarily $s_{2}=s_{1}$, so that from Eq.(%
\ref{bis}) $s_{1}$ is actually given by%
\begin{equation}
s_{1}=ct^{\prime }=s_{2}.  \label{s1t}
\end{equation}%
As a result, the integral $S_{C}^{(self)}$ reduces to%
\begin{equation}
S_{C}^{(self)}(r^{\prime },\left[ r^{\prime }\right] )=\frac{q}{4\pi \sigma
^{2}c}\int_{1}^{2}dx^{\prime }dy^{\prime }dz^{\prime }A^{^{\prime }(self)\mu
}(r^{\prime })\int_{-\infty }^{+\infty }dsu^{\prime \mu }(s)\delta
(\left\vert x^{\prime }\left( s\right) \right\vert -\sigma ),
\label{AAAAA-INTEGRAL}
\end{equation}%
with $x^{\prime \mu }\left( s\right) =r^{\prime \mu }-r^{\prime \mu }\left(
s\right) $. Moreover%
\begin{equation}
A_{\mu }^{\prime (self)}(r^{\prime })=2q\int_{-\infty }^{+\infty }ds^{\prime
\prime }u_{\mu }^{\prime }\left( s^{\prime \prime }\right) \delta (\widehat{R%
}^{\prime \alpha }\widehat{R}_{\alpha }^{\prime }-\rho ^{\prime 2}),
\end{equation}%
with $\widehat{R}^{\prime \alpha }=r^{\prime \alpha }-r^{\prime \alpha
}(s^{\prime \prime })$ and, thanks to Lemma 1,%
\begin{equation}
\rho ^{\prime 2}=\left\{
\begin{array}{ccc}
0 &  & (X^{\prime \alpha }X_{\alpha }^{\prime }\leq -\sigma ^{2}), \\
\rho ^{\prime 2}\equiv \sigma ^{2}\left[ 1+\frac{X^{\prime \alpha }X_{\alpha
}^{\prime }}{\sigma ^{2}}\right] &  & (X^{\prime \alpha }X_{\alpha }^{\prime
}>-\sigma ^{2}).%
\end{array}%
\right.
\end{equation}

Notice here that in $S_{C}^{(self)}(r^{\prime },\left[ r^{\prime }\right] )$
the contributions of the external and internal domains for the self-field
can be explicitly taken into account letting%
\begin{eqnarray}
\delta (\widehat{R}^{\prime \alpha }\widehat{R}_{\alpha }^{\prime }-\rho
^{\prime 2}) &=&\Theta (\sigma ^{2}+\xi ^{\alpha }\xi _{\alpha })\delta (%
\widehat{R}^{\prime \alpha }\widehat{R}_{\alpha }^{\prime }-\sigma
^{2}-X^{\prime \alpha }X_{\alpha }^{\prime })+  \notag \\
&&+\widehat{\Theta }(-\xi ^{\alpha }\xi _{\alpha }-\sigma ^{2})\delta (%
\widehat{R}^{\prime \alpha }\widehat{R}_{\alpha }^{\prime }),
\end{eqnarray}%
with $\Theta (x)$ and $\widehat{\Theta }(x)$ denoting respectively the
strong and weak Heaviside step functions
\begin{eqnarray}
\widehat{\Theta }(x) &=&\left\{
\begin{array}{lll}
1 &  & x\geq 0 \\
0 &  & x<0%
\end{array}%
\right. \\
\Theta (x) &=&\left\{
\begin{array}{lll}
1 &  & x>0 \\
0 &  & x\leq 0.%
\end{array}%
\right.
\end{eqnarray}%
On the other hand, the only contribution to the integral (\ref%
{AAAAA-INTEGRAL}) arises (because of the Dirac-delta in the current density)
from the subdomain for which $-\xi ^{\alpha }\xi _{\alpha }-\sigma ^{2}=0$.
Hence, $S_{C}^{(self)}$ simply reduces to the functional form:%
\begin{eqnarray}
S_{C}^{(self)}(r^{\prime },\left[ r^{\prime }\right] ) &=&\frac{2q^{2}}{4\pi
\sigma ^{2}c}\int_{0}^{\pi }d\vartheta ^{\prime }\sin \vartheta ^{\prime
}\int_{0}^{2\pi }d\varphi ^{\prime }\int_{0}^{+\infty }d\rho ^{\prime }\rho
^{^{\prime }2}\times  \notag \\
&&\times \int_{-\infty }^{+\infty }ds^{\prime \prime }u_{\mu }^{\prime
}\left( s^{\prime \prime }\right) \delta (\widehat{R}^{\prime \alpha }%
\widehat{R}_{\alpha }^{\prime })\int_{-\infty }^{+\infty }dsu^{\prime \mu
}(s)\delta (\left\vert x^{\prime }\left( s\right) \right\vert -\sigma ).
\end{eqnarray}%
The remaining spatial integration can now be performed letting%
\begin{equation}
\rho ^{\prime }\equiv \left\vert x^{\prime }\left( s\right) \right\vert
\end{equation}%
and making use of the spherical symmetry of the charge distribution. The
constraints placed by the two Dirac-delta functions $\delta (\widehat{R}%
^{\prime \alpha }\widehat{R}_{\alpha }^{\prime })$ and $\delta (\left\vert
x^{\prime }\left( s\right) \right\vert -\sigma )$ in the previous equation
imply that both $\widehat{R}^{\prime \alpha }\widehat{R}_{\alpha }^{\prime }$
and $\left\vert x^{\prime }\left( s\right) \right\vert $ are 4-scalars.
Then, introducing the representation%
\begin{equation}
\widehat{R}^{\prime \alpha }\equiv r^{\prime \alpha }-r^{\prime \alpha
}(s^{\prime \prime })=\widetilde{R}^{\prime \alpha }+x^{\prime \alpha
}\left( s\right) ,
\end{equation}%
with%
\begin{eqnarray}
\widetilde{R}^{\prime \alpha } &\equiv &r^{\prime \alpha }\left( s\right)
-r^{\prime \alpha }(s^{\prime \prime }), \\
x^{\prime \alpha }\left( s\right) &\equiv &r^{\prime \alpha }-r^{\prime
\alpha }\left( s\right) ,
\end{eqnarray}%
it follows that%
\begin{equation}
\widehat{R}^{\prime \alpha }\widehat{R}_{\alpha }^{\prime }=\widetilde{R}%
^{\prime \alpha }\widetilde{R}_{\alpha }^{\prime }+x^{\prime \alpha }\left(
s\right) x_{\alpha }^{\prime }\left( s\right) +2\widetilde{R}^{\prime \alpha
}x_{\alpha }^{\prime }\left( s\right)
\end{equation}%
is necessarily a 4-scalar independent of the integration angles $\left(
\varphi ^{\prime },\vartheta ^{\prime }\right) $ when evaluated on the
hypersurface $\Sigma :\widehat{R}^{\prime \alpha }\widehat{R}_{\alpha
}^{\prime }=0$. Similarly, the Dirac-delta $\delta (\left\vert x^{\prime
}\left( s\right) \right\vert -\sigma )$ warrants that %
$x^{\prime \alpha }\left( s\right) x_{\alpha }^{\prime }\left( s\right)
=-\sigma ^{2},$ 
which is manifestly a 4-scalar too. Let us now prove that necessarily
\begin{equation}
\widetilde{R}^{\prime \alpha }x_{\alpha }^{\prime }\left( s\right) \equiv 0.
\label{sym0}
\end{equation}%
In fact, on $\Sigma $ it must be%
\begin{eqnarray}
\frac{d}{ds}\left[ \widehat{R}^{\prime \alpha }\widehat{R}_{\alpha }^{\prime
}\right] &=&\frac{d}{ds^{\prime \prime }}\left[ \widehat{R}^{\prime \alpha }%
\widehat{R}_{\alpha }^{\prime }\right] =0,  \label{sym1} \\
\frac{d}{ds}\left[ \widetilde{R}^{\prime \alpha }x_{\alpha }^{\prime }\left(
s\right) \right] &=&u^{\prime \alpha }\left( s\right) x_{\alpha }^{\prime
}\left( s\right) -\widetilde{R}^{\prime \alpha }u_{\alpha }^{\prime }\left(
s\right) =-\widetilde{R}^{\prime \alpha }u_{\alpha }^{\prime }\left(
s\right) =-\frac{1}{2}\frac{d}{ds}\left[ \widetilde{R}^{\prime \alpha }%
\widetilde{R}_{\alpha }^{\prime }\right] , \\
\frac{d}{ds^{\prime \prime }}\left[ \widetilde{R}^{\prime \alpha }x_{\alpha
}^{\prime }\left( s\right) \right] &=&-u^{\prime \alpha }\left( s^{\prime
\prime }\right) x_{\alpha }^{\prime }\left( s\right) , \\
\frac{d}{ds^{\prime \prime }}\left[ \widetilde{R}^{\prime \alpha }\widetilde{%
R}_{\alpha }^{\prime }\right] &=&-2\widetilde{R}^{\prime \alpha }u_{\alpha
}^{\prime }\left( s^{\prime \prime }\right) .
\end{eqnarray}%
Therefore,%
\begin{equation}
\frac{d}{ds}\left[ \widehat{R}^{\prime \alpha }\widehat{R}_{\alpha }^{\prime
}\right] =\frac{d}{ds}\left[ \widetilde{R}^{\prime \alpha }\widetilde{R}%
_{\alpha }^{\prime }+2\widetilde{R}^{\prime \alpha }x_{\alpha }^{\prime
}\left( s\right) \right] =0,
\end{equation}%
\begin{eqnarray}
\frac{d}{ds^{\prime \prime }}\left[ \widehat{R}^{\prime \alpha }\widehat{R}%
_{\alpha }^{\prime }\right] &=&\frac{d}{ds^{\prime \prime }}\left[
\widetilde{R}^{\prime \alpha }\widetilde{R}_{\alpha }^{\prime }+2\widetilde{R%
}^{\prime \alpha }x_{\alpha }^{\prime }\left( s\right) \right] =  \notag \\
&=&-2\widetilde{R}^{\prime \alpha }u_{\alpha }^{\prime }\left( s^{\prime
\prime }\right) -2u^{\prime \alpha }\left( s^{\prime \prime }\right)
x_{\alpha }^{\prime }\left( s\right) =0,
\end{eqnarray}%
from which it follows that, on $\Sigma $, $\widetilde{R}^{\prime \alpha }$
is a 4-vector, since by definition both $u_{\alpha }^{\prime }\left(
s^{\prime \prime }\right) $ and $x_{\alpha }^{\prime }\left( s\right) $ are
4-vectors too. Now we notice that%
\begin{equation}
\widetilde{R}^{\prime \alpha }\widetilde{R}_{\alpha }^{\prime }=f\left(
s,s^{\prime \prime }\right) =f\left( s^{\prime \prime },s\right) ,
\label{sym2}
\end{equation}%
with $f$ being a 4-scalar which is symmetric with respect to $s$ and $%
s^{\prime \prime }$, while by construction%
\begin{equation}
\widetilde{R}^{\prime \alpha }x_{\alpha }^{\prime }\left( s\right) =g\left(
s,s^{\prime \prime },\sigma \right) \neq g\left( s^{\prime \prime },s,\sigma
\right) ,  \label{sym3}
\end{equation}%
where $g$ is a non-symmetric 4-scalar with respect to the same parameters.
On the other hand, Eq.(\ref{sym1}) requires that $\widehat{R}^{\prime \alpha
}\widehat{R}_{\alpha }^{\prime }$ must be symmetric in both $s$ and $%
s^{\prime \prime }$, so that, thanks to Eqs.(\ref{sym2}) and (\ref{sym3}),
we can conclude that $g=g\left( \sigma \right) $ is a constant 4-scalar
which can depend at most on $\sigma $. To determine the precise value of $g=%
\widetilde{R}^{\prime \alpha }x_{\alpha }^{\prime }\left( s\right) $ we
evaluate it in the COS comoving reference frame, where by definition $%
r_{COS}^{\mu }\left( s_{0}\right) =\left( s_{0},\mathbf{0}\right) $ for all
the COS proper times $s_{0}\in \lbrack -\infty ,+\infty ]$. In this frame $%
\widetilde{R}^{\prime \alpha }=\left( s-s^{\prime \prime },\mathbf{0}\right)
$ has only time component and when $s_{0}=s$ we get $g=\widetilde{R}^{\prime
\alpha }x_{\alpha }^{\prime }\left( s\right) =0$ identically. On the other
hand, since $g$ is a 4-scalar, it is independent of both $s$ and $s^{\prime
\prime }$ and it is null when $s_{0}=s$, we conclude that it must be null
for all $s_{0}$ and in any reference frame, which proves Eq.(\ref{sym0}).

Hence, as a result of the integration, the action integral $S_{C}^{(self)}$
takes necessarily the expression%
\begin{equation}
S_{C}^{(self)}(r^{\prime },\left[ r^{\prime }\right] )=\frac{2q^{2}}{c}%
\int_{1}^{2}dr_{\mu }^{\prime }\left( s^{\prime \prime }\right)
\int_{1}^{2}dr^{\prime \mu }(s^{\prime })\delta (\widetilde{R}^{\prime
\alpha }\widetilde{R}_{\alpha }^{\prime }-\sigma ^{2}).
\end{equation}%
Finally, since by construction $S_{C}^{(self)}$ is a 4-scalar, it follows
that the primes can be dropped thus yelding the following representation
holding in a general reference frame:%
\begin{equation}
S_{C}^{(self)}(r,\left[ r\right] )=\frac{2q^{2}}{c}\int_{1}^{2}dr_{\mu
}\left( s\right) \int_{1}^{2}dr^{\mu }(s^{\prime })\delta (\widetilde{R}%
^{\alpha }\widetilde{R}_{\alpha }-\sigma ^{2}),
\end{equation}%
where for simplicity of notation $s^{\prime \prime }$ has been replaced with
$s^{\prime }$ and $\widetilde{R}^{\alpha }$ now denotes%
\begin{equation}
\widetilde{R}^{\alpha }\equiv r^{\alpha }\left( s\right) -r^{\alpha
}(s^{\prime }).
\end{equation}
It is worth pointing out the following basic properties of the functional $%
S_{C}^{(self)}$:

1) it is a non-local functional in the sense that it contains a coupling
between the past and the future of the dynamical system (see Eq.(\ref{dynsys}%
)). In fact it can be equivalently represented as%
\begin{equation}
S_{C}^{(self)}(r,\left[ r\right] )=\frac{2q^{2}}{c}\int_{-\infty }^{+\infty
}ds\frac{dr_{\mu }\left( s\right) }{ds}\int_{-\infty }^{+\infty }ds^{\prime }%
\frac{dr^{\mu }(s^{\prime })}{ds^{\prime }}\delta (\widetilde{R}^{\alpha }%
\widetilde{R}_{\alpha }-\sigma ^{2});
\end{equation}%
2) furthermore, it is symmetric, namely it fulfills the property%
\begin{equation}
S_{C}^{(self)}(r_{A},\left[ r_{B}\right] )=S_{C}^{(self)}(r_{B},\left[ r_{A}%
\right] ),
\end{equation}%
where $r_{A}$ and $r_{B}$ are two arbitrary curves of the functional class $%
\left\{ f\right\} $ (see Eq.(\ref{FUNCTIONAL CLASS})). \bigskip

\subsection{$S_{C}^{\left( ext\right) }(r)$: EM coupling with the external
field}

The action integral $S_{C}^{\left( ext\right) }(r)$ of the EM coupling with
the external field is a 4-scalar defined as%
\begin{equation}
S_{C}^{\left( ext\right) }(r)=\int_{1}^{2}d\Omega \frac{1}{c^{2}}A^{(ext)\mu
}(r)j_{\mu }\left( r\right) ,
\end{equation}%
where $A^{(ext)\mu }(r)$ is the 4-vector potential of the external field,
assumed to be assigned, and $j_{\mu }\left( r\right) $ is the current
density given by Eq.(\ref{intj}). The evaluation of the action integral $%
S_{C}^{\left( ext\right) }$ proceeds exactly in the same way as outlined for
$S_{C}^{\left( self\right) },$ with the introduction of the Lorentz boost (%
\ref{Lboost}), the spherical spatial coordinates and the use of the result
from Lemma 3. The only difference now is that the vector potential $%
A^{(ext)\mu }(r)$ does not possess spherical symmetry when evaluated in $%
S_{NI}$. As a result, spatial integration over the angle variables $%
\vartheta ^{\prime }$ and $\varphi ^{\prime }$ cannot be computed
explicitly. This leads to the introduction of the \textit{surface average EM
external 4-potential }$\overline{A}^{(ext)\mu },$ which is defined in $%
S_{NI} $ as%
\begin{equation}
\overline{A}^{\prime (ext)\mu }\left( r^{\prime }\left( s\right) ,|x^{\prime
}|\right) \equiv \frac{1}{4\pi }\int_{0}^{2\pi }d\varphi ^{\prime
}\int_{0}^{\pi }d\vartheta ^{\prime }\sin \vartheta ^{\prime }\left[
A^{^{\prime }(ext)\mu }(r^{\prime \mu }(s)+x^{\prime \mu })\right] ,
\label{surfavera}
\end{equation}%
where we have used the relation (\ref{xx}). With this definition, the time
and radial integrals can then be calculated using the Dirac-delta functions
as outlined for the self-coupling action integral. After performing a final
transformation to an arbitrary reference frame, this gives the following
expression for $S_{C}^{\left( ext\right) }$:%
\begin{equation}
S_{C}^{(ext)}(r)=\frac{q}{c}\int_{1}^{2}\overline{A}^{(ext)\mu }\left(
r^{\mu }\left( s\right) ,\sigma \right) dr_{\mu }(s).
\end{equation}

\bigskip

\subsection{$S_{\protect\chi }(u,\protect\chi )$: kinematic constraint}

The kinematic constraint concerns the normalization of the extremal
4-velocity of the COS. This is defined as%
\begin{equation}
S_{\chi }(u,\chi )\equiv \int_{-\infty }^{+\infty }ds\chi (s)\left[ u_{\mu
}(s)u^{\mu }(s)-1\right] ,  \label{const}
\end{equation}%
where $\chi (s)$ is a Lagrange multiplier.

\bigskip

\subsection{$S_{M}(r,u)$: inertial mass functional}

The action integral $S_{M}$ of the inertial mass for the extended particle
is here defined as the following 4-scalar:%
\begin{equation}
S_{M}(r,u)\equiv \int_{1}^{2}d\Omega \frac{1}{c}g_{\mu \nu }T_{M}^{\mu \nu
}\left( r\right) ,  \label{sm1}
\end{equation}%
where $d\Omega $ denotes the invariant 4-volume element and $T_{M}^{\mu \nu
} $ the stress-energy tensor corresponding to the mass distribution of the
finite-size charged particle. Notice that the choice of $S_{M}$ is
consistent with the customary definition of the stress-energy tensor $T^{\mu
\nu }$ (for a fluid or a field) in terms of $T^{\mu \nu }\equiv \frac{\delta
L}{\delta g_{\mu \nu }}$, with $L$ being a suitable Lagrangian function and $%
\delta $ representing the variational derivative \cite{LL}. Therefore, it is
natural to identify $S_{M}$ with the trace of the mass stress-energy tensor
for the extended particle. In particular, the explicit representation of $%
T_{M}^{\mu \nu }$ follows by projecting the mass current density $%
j_{mass}^{\mu }(r)$ given in Eq.(\ref{mass_current}) along the velocity of a
generic shell mass-element parameterized in terms of the proper time $s$ of
the COS. The procedure is completely analogous to that outlined in Section
3. Equivalently, $T_{M}^{\mu \nu }$ can also be derived by considering the
stress-energy tensor of a perfect fluid without pressure (since the mass is
located on a shell by assumption) and imposing the rigidity constraints (\ref%
{eee1}) and (\ref{eee2}). Accordingly, one obtains the following expression:%
\begin{equation}
T_{M}^{\mu \nu }\left( r\right) \equiv \frac{m_{o}c^{2}}{4\pi \sigma ^{2}}%
\int_{-\infty }^{+\infty }dsu^{\mu }(s)u^{\nu }\left( s\right) \left[
1+\Gamma \right] \delta (\left\vert x\right\vert -\sigma )\delta (s-s_{1}),
\label{tmunu}
\end{equation}%
with $s_{1}$ being the root of Eq.(\ref{s1}) and $\Gamma $ the contribution
of the Thomas precession. We notice that the stress-energy tensor thus
defined is symmetric. With this definition, the action integral $S_{M}$
becomes%
\begin{equation}
S_{M}(r,u)\equiv \frac{m_{o}c}{4\pi \sigma ^{2}}\int_{1}^{2}d\Omega
\int_{-\infty }^{+\infty }dsu^{\mu }(s)u_{\mu }\left( s\right) \left[
1+\Gamma \right] \delta (\left\vert x\right\vert -\sigma )\delta (s-s_{1}).
\end{equation}%
The integration over the 4-volume element can be performed explicitly in the
same way as explained before (for the EM coupling action integral), by using
the prescription of Lemma 3 and the transformation to local spatial
spherical coordinates. In particular, here we notice that both $u^{\mu }(s)$
and $\frac{du_{\alpha }}{ds}$ appearing in $\Gamma $ are independent of the
integration variables, while in the reference system $S_{NI}$ introduced in
Lemma 3 we have that %
$\int_{0}^{\pi }d\vartheta ^{\prime }\sin \vartheta ^{\prime }\int_{0}^{2\pi
}d\varphi ^{\prime }\xi ^{\alpha }=0,$ 
as it follows from the property of $\xi ^{\alpha }$ to be a pure spatial
vector in $S_{NI}$. Thanks to this feature, the whole integral is
straightforward, so that one obtains for $S_{M}(r,u)$ the final expression:%
\begin{equation}
S_{M}(r,u)\equiv \int_{1}^{2}m_{o}cu_{\mu }dr^{\mu }  \label{sm2}
\end{equation}%
holding in an arbitrary reference frame. Concerning the solution (\ref{sm2}%
), a remark is in order. The choice of $S_{M}$ given by Eq.(\ref{sm1})
proves to be the correct one. In fact, as expected Eq.(\ref{sm2}) is
formally the same action integral of a point particle, with the difference
that here $u_{\mu }$ represents the 4-velocity of the COS rather than the
one of a point mass.

\bigskip

\section{The variational Lagrangian}

In this section we collect together all of the contributions to $S_{1}$
previously obtained. From the results of the previous section we can write
the action integral $S_{1}$ as a line integral in terms of a variational
Lagrangian $L_{1}(r,\left[ r\right] ,u,\chi )$ as follows [see Eq.(\ref%
{HAMILTON ACTION})]:%
\begin{equation}
S_{1}=\int_{-\infty }^{+\infty }dsL_{1}(r,\left[ r\right] ,u,\chi ).
\end{equation}%
More precisely, $L_{1}(r,\left[ r\right] ,u,\chi )$ is defined as:%
\begin{equation}
L_{1}(r,\left[ r\right] ,u,\chi )=L_{M}(r,u)+L_{\chi }(u,\chi
)+L_{C}^{(ext)}(r)+L_{C}^{(self)}(r,\left[ r\right] ),
\label{EXTREMAL LAGRANGIAN}
\end{equation}%
where%
\begin{eqnarray}
L_{M}(r,u) &=&m_{o}cu_{\mu }\frac{dr^{\mu }}{ds},  \label{LAGRANGIAN -MASS}
\\
L_{\chi }(u,\chi ) &=&\chi (s)\left[ u_{\mu }(s)u^{\mu }(s)-1\right] ,
\label{LAGRANGIAN -constarint-2} \\
L_{C}^{(ext)}(r) &=&\frac{dr}{ds}^{\mu }\frac{q}{c}\overline{A}_{\mu
}^{(ext)}(r(s),\sigma ),  \label{LAGRANGIAN -EXTERNAL EM}
\end{eqnarray}%
denote the local contributions respectively from the inertial, the
constraint and the external EM field coupling terms, while
\begin{equation}
L_{C}^{(self)}(r,\left[ r\right] )=\frac{2q^{2}}{c}\frac{dr}{ds}^{\mu
}\int_{1}^{2}dr_{\mu }^{\prime }\delta (\widetilde{R}^{\mu }\widetilde{R}%
_{\mu }-\sigma ^{2})  \label{LAGRANGIAN-SELF-EM}
\end{equation}%
represents the non-local contribution arising from the EM self-field
coupling.

The conclusion is remarkable. Indeed, although the extended particle can be
regarded as a continuous system carrying mass and charge current densities,
the variational functional here determined is similar to that of a point
particle subject to appropriate interactions. In fact, because of the
rigidity constraint and the spherical symmetry imposed on the charge and
mass distributions, the variational action $S_{1}$ is actually reduced from
a volume integral to a line integral over the proper time of the COS. This
is realized by means of the volume integration performed in the reference
frame $S_{NI}$ and thanks to Lemma 3.

The procedure introduces the surface-average operator acting both on the
external and the self EM coupling terms. As a result, the Lagrangian (\ref%
{EXTREMAL LAGRANGIAN}) must be interpreted as prescribing the dynamics for
the COS of the charged particle in terms of averaged EM fields, integrating
all the force contributions to the translational motion on the shell.
Furthermore, we recall once again the formal analogy between the Lagrangian $%
L_{M}(r,u)$ and the one of a point particle, when $u_{\mu }$ is interpreted
as the 4-velocity of the point mass rather than that of the COS of the
shell. This means that the dynamics of the finite-size particle is
effectively described in terms of a point particle with a finite-size charge
distribution. Hence, \emph{the mathematical problem is formally the same of
that for a Lorentzian particle}. Therefore, this proves that the particular
case of a Lorentzian particle is formally included in the present
description, in the limit in which the radius of the mass distribution $%
\sigma _{m}$ is sent to zero while keeping the charge spatial extension
fixed ($\sigma >0$). The conclusion manifestly follows within the framework
of special relativity, in which any possible curvature effects due to the EM
field and the mass of the particle itself are neglected.

\section{The variational principle and the RR equation}

In this section we shall determine the explicit form of the relativistic RR
equation for the non-rotating charged particle. As pointed out earlier \cite%
{Tessarotto2008c}, this goal can be uniquely attained by means of a \textit{%
synchronous variational principle}, in analogy with the approach originally
developed for point particles by Nodvik in terms of an asynchronous
principle (Nodvik, 1964 \cite{Nodvik1964}). In particular, we intend to
prove that, in the present case, the exact RR equation can be \emph{uniquely}
and \emph{explicitly} obtained by using the hybrid synchronous Hamilton
variational principle defined in the previous section and given by Eq.(\ref%
{pointp}). In this case the action functional is expressed by means of
superabundant hybrid (i.e., non-Lagrangian) variables and the variations are
considered as synchronous, i.e., they are performed by keeping constant the
particle COS proper time. Taking into account the results presented in the
previous sections, the appropriate form of the Hamilton variational
principle is given by the following theorem:\newline

\textbf{THM.1 - Hybrid synchronous Hamilton variational principle}

\emph{In validity of the SR-CE axioms, let us assume that: }

\begin{enumerate}
\item \emph{the Hamilton action }$S_{1}(r,u,\chi ,\left[ r\right] )$\emph{\
is defined by Eq.(\ref{pointp}),} \emph{with }$A_{\mu }^{(self)}$\emph{\
given by Eq.(\ref{AAAAA=INTEGRAL REPRESENTATION}) and }$\chi (s)$ \emph{%
being a suitable Lagrange multiplier;}

\item \emph{the real functions }$f(s)$ \emph{in the functional class }$%
\left\{ f\right\} $ \emph{[see Eq.(\ref{FUNCTIONAL CLASS})]}\emph{\ are
identified with}%
\begin{equation}
\emph{\ }f(s)\equiv \left[ r^{\mu }(s),u_{\mu }(s),\chi (s)\right] ,
\label{variational functions}
\end{equation}%
\emph{with synchronous variations }$\delta f(s)\equiv $\emph{\ }$f(s)-$\emph{%
\ }$f_{1}(s)$\emph{\ belonging to}%
\begin{eqnarray}
\left\{ \delta f\right\} &\equiv& \delta f_{i}(s):\delta
f_{i}(s)=f_{i}(s)-f_{1i}(s);  \notag \\
&&\text{ }i=1,n \emph{\ and } \forall f(s),f_{1}(s)\in \left\{ f\right\},
\label{SYNCHRONOUS FUNCTIONAL CLASS}
\end{eqnarray}%
\emph{here referred to as the functional class of synchronous variations;\ }

\item \emph{the extremal curve }$f\in \left\{ f\right\} $ \emph{of }$S_{1},$
\emph{which is the solution of the equation}%
\begin{equation}
\delta S_{1}(r,u,\chi ,\left[ r\right] )=0,
\label{HYBRID HAMILTON
VARIATIONAL PRINCIPLE}
\end{equation}%
\emph{exists for arbitrary variations} $\delta f(s)$ \emph{(hybrid
synchronous Hamilton variational principle);}

\item \emph{if }$r^{\mu }(s)$ \emph{is extremal, the line element }$ds$\emph{%
\ satisfies the constraint }$ds^{2}=\eta _{\mu \nu }dr^{\mu }(s)dr^{\nu }(s)$%
\emph{;}

\item \emph{the 4-vector field }$A_{\mu }^{(ext)}(r)$ \emph{is suitably
smooth in the whole Minkowski space-time }$M^{4}$\emph{;}

\item \emph{the E-L equation for the extremal curve }$r^{\mu }(s)$\emph{\ is
determined subject to the constraint that the delay-time $s_{ret}$ (namely
the root of the delay-time equation (\ref{delayshell}) below)}\emph{\ must
be chosen consistently with ECP.}
\end{enumerate}

\emph{It then follows that:} 

\emph{T1}$_{1})$\emph{\ If all the synchronous variations }$\delta f_{i}(s)$
\emph{(}$\emph{i=1,n}$\emph{) are considered as being independent, the E-L
equations for $\chi (s)$ and $u_{\mu }$ following from the synchronous
hybrid Hamilton variational principle (\ref{HYBRID HAMILTON VARIATIONAL
PRINCIPLE})} \emph{give respectively}
\begin{eqnarray}
&&\left. \frac{\delta S_{1}}{\delta \chi (s)}=u_{\mu }u^{\mu }-1=0,\right.
\label{Eq.-------------------32} \\
&&\left. \frac{\delta S_{1}}{\delta u_{\mu }}=m_{o}cdr^{\mu }+2\chi u^{\mu
}ds=0.\right.  \label{Eq.-------------------32A}
\end{eqnarray}%
\emph{Instead, the E-L equation for $r_{\mu }$}%
\begin{equation}
\frac{\delta S_{1}}{\delta r^{\mu }(s)}=0
\label{Eq...............................32B}
\end{equation}%
\emph{yields the following covariant (and hence also MLC)\ 4-vector,
second-order delay-type ODE:}
\begin{equation}
m_{o}c\frac{du_{\mu }(s)}{ds}=\frac{q}{c}\overline{F}_{\mu \nu
}^{(ext)}(r(s))\frac{dr^{\nu }(s)}{ds}+\frac{q}{c}\overline{F}_{\mu
k}^{\left( self\right) }\left( r\left( s\right) ,r\left( s^{\prime }\right)
\right) \frac{dr^{k}(s)}{ds},  \label{Eq.---------------20B}
\end{equation}%
\emph{which is identified with the RR equation of motion for the COS of a
spherical shell non-rotating charge particle. Here }
\begin{equation}
\overline{F}_{\mu \nu }^{(ext)}\equiv \partial _{\mu }\overline{A}_{\nu
}^{(ext)}-\partial _{\nu }\overline{A}_{\mu }^{(ext)}
\label{EXTERNAL EM FIELD}
\end{equation}%
\emph{denotes the surface-average [defined according to Eq.(\ref{surfavera}%
)] of the Faraday tensor carried by the externally-generated EM 4-vector and
evaluated at the particle 4-position }$r^{\mu }(s).$ \emph{In addition, }$%
\overline{F}_{\mu k}^{\left( self\right) }$\emph{\ - in MLC 4-vector
representation - is the surface-averaged Faraday tensor of the corresponding
EM self-field, given by }%
\begin{equation}
\overline{F}_{\mu k}^{\left( self\right) }=-\frac{2q}{\left\vert \widetilde{R%
}^{\alpha }u_{\alpha }(s^{\prime })\right\vert }\frac{d}{ds^{\prime }}%
\left\{ \frac{u_{\mu }(s^{\prime })\widetilde{R}_{k}-u_{k}(s^{\prime })%
\widetilde{R}_{\mu }}{\widetilde{R}^{\alpha }u_{\alpha }(s^{\prime })}%
\right\} _{s^{\prime }=s-s_{ret}}.  \label{COVARIANT FORM}
\end{equation}%
\emph{Imposing the constraint }$ds^{\prime }=\gamma \left( t^{\prime
}\right) cdt^{\prime }$\emph{, this implies also}%
\begin{eqnarray}
\overline{F}_{\mu k}^{\left( self\right) } &=&-\frac{2q}{c\left\vert
(t-t^{\prime })-\frac{1}{c^{2}}\frac{d\mathbf{r}(t^{\prime })}{dt^{\prime }}%
\cdot (\mathbf{r-r}\left( t^{\prime }\right) )\right\vert }  \notag \\
&&\frac{d}{dt^{\prime }}\left\{ \frac{v_{\mu }(t^{\prime })\widetilde{R}%
_{k}-v_{k}(t^{\prime })\widetilde{R}_{\mu }}{c^{2}\left[ (t-t^{\prime })-%
\frac{1}{c^{2}}\frac{d\mathbf{r}(t^{\prime })}{dt^{\prime }}\cdot (\mathbf{%
r-r}\left( t^{\prime }\right) )\right] }\right\} _{t^{\prime }=t-t_{ret}}.
\label{A_NU_K2}
\end{eqnarray}%
\emph{Here }$u^{\mu }=\frac{dr^{\mu }}{ds}$\emph{\ denotes the COS
4-velocity and }$v^{\mu }(t)=\frac{dr^{\mu }}{dt}$\emph{, while }$%
s_{ret}=s-s^{\prime }$\emph{\ is the positive root of the delay-time equation%
}
\begin{equation}
\widetilde{R}^{\alpha }\widetilde{R}_{\alpha }-\sigma ^{2}=0.
\label{delayshell}
\end{equation}

\emph{T1}$_{2})$\emph{\ The E-L equations (\ref{Eq.-------------------32}),(%
\ref{Eq.-------------------32A}) and (\ref%
{Eq...............................32B})} \emph{imply that the extremal
functional takes the form}%
\begin{equation}
S(r,\left[ r\right] ,u,)=S_{1}(r,\left[ r\right] ,u,\chi (s)=-\frac{m_{o}c}{2%
}).  \label{extremal functional}
\end{equation}

\emph{T1}$_{3})$ \emph{If }$F_{\mu }^{(ext)\nu }(r)\equiv 0$ \emph{for all }$%
s\leq s_{1}\in \mathbb{R},$ \emph{a particular solution of Eq.(\ref%
{Eq.---------------20B}), holding for all }$s\leq s_{1}$ \emph{is provided
by the inertial motion, i.e.,}%
\begin{eqnarray}
\frac{dr^{\mu }(s)}{ds} &=&u_{o}^{\mu }=const.,  \label{INERTIAL-1} \\
\frac{du^{\mu }}{ds} &=&0,  \label{INERTIAL-2}
\end{eqnarray}%
\emph{\ in agreement with the Galilei principle of inertia. }

\emph{T1}$_{4})$ \emph{The RR equation Eq.(\ref{Eq.---------------20B}) also
holds for a Lorentzian particle having the same charge distribution of the
finite-size particle }$(\sigma >0)$\emph{\ and carrying a point-mass with
position and velocity 4-vectors }$r^{\mu }(s),$ $u^{\mu }(s)$.

\emph{Proof} - \emph{T1}$_{1})$ and \emph{T1}$_{2})$ The proof proceeds as
follows. Since $\frac{\partial }{\partial u^{\mu }}\delta (\widetilde{R}%
^{\alpha }\widetilde{R}_{\alpha }-\sigma ^{2})=\frac{\partial }{\partial
u^{\prime \mu }}\delta (\widetilde{R}^{\alpha }\widetilde{R}_{\alpha
}-\sigma ^{2})\equiv 0$, the variations with respect to $\chi (s)$ and $%
u_{\mu }$ deliver respectively the two E-L equations (\ref%
{Eq.-------------------32}) and (\ref{Eq.-------------------32A}). Hence,
the Lagrange multiplier $\chi $ must be for consistency%
\begin{equation}
2\chi =-m_{o}c,
\end{equation}%
so that, ignoring gauge contributions with respect to $\chi $, the extremal
functional $S_{1}(r,u,\chi ,\left[ r\right] )$ takes the form (\ref{extremal
functional}) [statement \emph{T1}$_{2}$]. To prove also Eq.(\ref%
{Eq...............................32B}), we notice that the synchronous
variation of\ $S_{C}^{(self)}$ has the form
\begin{equation}
\delta S_{C}^{(self)}=\delta A+\delta B,  \label{Eq.---A.3}
\end{equation}%
where%
\begin{equation}
\begin{array}{c}
\delta A\equiv -\frac{4q^{2}}{c}\eta _{\mu \nu }\int_{1}^{2}\delta r^{\mu }d%
\left[ \int_{1}^{2}dr^{\prime \nu }\delta (\widetilde{R}^{\alpha }\widetilde{%
R}_{\alpha }-\sigma ^{2})\right] , \\
\delta B\equiv \frac{4q^{2}}{c}\eta _{\alpha \beta }\int_{1}^{2}dr^{\prime
\beta }\int_{1}^{2}dr^{\alpha }\delta r^{\mu }\frac{\partial }{\partial
r^{\mu }}\delta (\widetilde{R}^{k}\widetilde{R}_{k}-\sigma ^{2}),%
\end{array}
\label{EQ.---A.4}
\end{equation}%
and $r^{\prime \nu }\equiv r^{\nu }\left( s^{\prime }\right) $ and $r^{\nu
}\equiv r^{\nu }\left( s\right) $. Then we can write
\begin{eqnarray}
\delta A &\equiv &\frac{2q}{c}\eta _{\mu \nu }\int_{1}^{2}\delta r^{\mu
}dr^{k}\left[ B_{k}^{\nu }\right] _{t^{\prime }=t-t_{ret}},
\label{EQ.---A.5} \\
\delta B &\equiv &-\frac{2q}{c}\eta _{\alpha \beta }\int_{1}^{2}\delta
r^{\mu }dr^{\alpha }\left[ B_{\mu }^{\beta }\right] _{t^{\prime }=t-t_{ret}},
\notag
\end{eqnarray}%
where $B_{k}^{\nu }$\ is
\begin{equation}
B_{k}^{\nu }\equiv -\frac{q}{c\left\vert (t^{\prime }-t)-\frac{1}{c^{2}}%
\frac{d\mathbf{r}(t^{\prime })}{dt^{\prime }}\cdot (\mathbf{r}^{\prime }%
\mathbf{-r})\right\vert }\frac{d}{dt^{\prime }}\left\{ \frac{v^{\nu
}(t^{\prime })\widetilde{R}_{k}}{c^{2}\left[ (t^{\prime }-t)-\frac{1}{c^{2}}%
\frac{d\mathbf{r}(t^{\prime })}{dt^{\prime }}\cdot (\mathbf{r-r}^{\prime })%
\right] }\right\}  \label{A_nu_k}
\end{equation}%
(the details of the derivation of these identities are provided in Appendix
A). Finally, from the results given in Appendix A, the variation with
respect to $r^{\mu }$ yields%
\begin{equation}
\frac{\delta S_{1}}{\delta r^{\mu }}=-m_{o}cdu_{\mu }(s)+\frac{q}{c}dr^{k}%
\overline{F}_{\mu k}^{\left( self\right) }+\frac{q}{c}\left[ \partial _{\mu }%
\overline{A}_{\nu }^{(ext)}(r(s))-\partial _{\nu }\overline{A}_{\mu
}^{(ext)}(r(s))\right] dr^{\nu },  \label{Eq.-------------------32B}
\end{equation}%
where
\begin{equation}
\overline{F}_{\mu k}^{\left( self\right) }=2(B_{k\mu }-B_{\mu k}),
\end{equation}%
from which Eqs.(\ref{Eq.---------------20B})-(\ref{delayshell}) follow. This
yields the \emph{RR equation} being sought, i.e., \emph{the} \emph{exact
relativistic equation of motion for the translational dynamics of the COS of
a finite-size spherical shell charge particle subject to the simultaneous
action of a prescribed external EM field and of its EM self-field.}

\emph{T1}$_{3})$ The proof of Eqs.(\ref{INERTIAL-1})-(\ref{INERTIAL-2}) is
straightforward. In fact, let us assume that in the interval $\left[ -\infty
,s_{1}\right] $ the motion is inertial, namely that $\frac{d}{ds}u_{\mu
}\equiv 0$\emph{,}$\forall s$ $\in \left[ -\infty ,s_{1}\right] .$ This
implies that in $\left[ -\infty ,s_{1}\right] $ it must be $u_{\mu }\equiv
u_{0\mu },$ with $u_{0\mu }$ denoting a constant 4-vector velocity. It
follows that $\forall s,s^{\prime }\in $ $\left[ -\infty ,s_{1}\right] ,$ $%
r_{\mu }(s)=r_{\mu }(s^{\prime })+u_{0\mu }(s^{\prime })(s-s^{\prime })$ and
$R_{\mu }=u_{0\mu }(s)(s-s^{\prime }).$ Hence, by direct substitution in Eq.(%
\ref{A_NU_K2}) we get that $v_{\mu }(t^{\prime })\widetilde{R}%
_{k}-v_{k}(t^{\prime })\widetilde{R}_{\mu }=0$, which by consequence implies
also that $dr^{k}H_{\mu k}\equiv 0$ identically in this case.

\emph{T1}$_{4})$ The proof follows immediately from the definition of
Lorentzian particle given above by noting that in the context of SR the
variational particle Lagrangian $L_{1}(r,\left[ r\right] ,u,\chi )$ [see Eq.(%
\ref{EXTREMAL LAGRANGIAN})] formally coincides with that of a Lorentzian
particle characterized by a finite charge distribution [i.e., with $\sigma
>0],$ subject to the simultaneous action of the averaged external and EM
self-fields $\overline{F}_{\mu \nu }^{(ext)}$ and $\overline{F}_{\mu
k}^{\left( self\right) }.$

\textbf{Q.E.D.}\newline

We notice that, by assumption, the varied functions\emph{\ }$f(s)\equiv %
\left[ r^{\mu }(s),u_{\mu }(s),\chi (s)\right] $\emph{\ }are \emph{%
unconstrained}, namely they are solely subject to the requirement that end
points and boundary values are kept fixed. This implies that all of the 9
components of the variations $\delta f(s),$ namely $\delta r^{\mu
}(s),\delta u_{\mu }(s),\delta \chi (s),$ must be considered independent. On
the other hand, the extremal curves $f(s)$ of $S_{1}(r,u,\chi ,\left[ r%
\right] ),$ the solution of the hybrid Hamilton variational principle,
satisfy all of the required physical constraints, so that only 6 of them are
actually independent. In fact, the resulting E-L equations determine,
besides the RR equation (\ref{Eq.---------------20B}), also the relationship
between $r^{\mu }(s)$ and $u_{\mu }(s)$, namely%
\begin{equation}
\left. u^{\mu }(s)=\frac{dr^{\mu }(s)}{ds},\right.
\end{equation}%
as well as the physical constraint%
\begin{equation}
\left. u^{\mu }(s)u_{\mu }(s)=1.\right.  \label{4-velocity
constraint}
\end{equation}%
As a consequence, $r^{\mu }(s)$ and $u^{\mu }(s)$ coincide respectively with
the physical 4-position and 4-velocity of the COS mass particle. Therefore
only 3 components of the 4-velocity are actually independent, while the
first component of the 4-position $ct$ can always be represented in terms of
the proper length $s$ (so that only the spatial part of the position
4-vector actually defines a set of independent Lagrangian coordinates)$.$

A further basic feature of the RR equation concerns the validity of GIP and
its meaning in this context. In fact, let us assume that the external EM
field is non-vanishing in the time interval $I_{12}\equiv \left[ s_{1},s_{2}%
\right] ,$ while it vanishes identically in $I_{2}\equiv \left[
s_{2},+\infty \right] .$ Then, the inertial solution (\ref{INERTIAL-1}) and (%
\ref{INERTIAL-2}) does not hold, by definition, in $I_{12}$ and is only
achieved in an asymptotic sense in $I_{2}$, i.e., in the limit $s\rightarrow
+\infty .$ In fact, the non-local feature of the RR effect prevents the
particle from reaching the inertial state in a finite time interval. It is
concluded, therefore, that GIP must be intended as holding \emph{in the past,%
} namely in the time interval\emph{\ }$s\leq s_{1}\in \mathbb{R},$ where by
assumption no external EM field is acting on the particle.

\bigskip

\section{Standard Lagrangian and conservative forms of the RR equation}

In this section we discuss some developments about the physical properties
of the non-local RR equation obtained, which exactly describes the
translational dynamics of the COS of a spherical-shell non-rotating charged
particle. Remarkably, the variational principle (THM.1) implies that the E-L
equations (\ref{Eq.-------------------32})-(\ref{Eq.---------------20B}) can
be cast in an equivalent way either:

1) in a standard Lagrangian form, namely expressed in the form of Lagrange
equations defined in terms of a suitable non-local effective Lagrangian $%
L_{eff};$

2) in a conservative form, as the divergence of a suitable effective
stress-energy tensor.

The result is provided by the following theorem.\newline

\textbf{THM.2 - RR equation in standard Lagrangian and conservative forms}

\emph{Given validity of THM.1, it follows that:}

\emph{T2}$_{1})$ \emph{Introducing the non-local real function }%
\begin{equation}
L_{eff}\equiv L_{M}(r,u)+L_{\chi }(u,\chi
)+L_{C}^{(ext)}(r)+2L_{C}^{(self)}(r,\left[ r\right] ),
\label{extremal Lagrangian-0}
\end{equation}%
\emph{here referred to as non-local effective Lagrangian, the E-L equations (%
\ref{Eq.-------------------32}),(\ref{Eq.-------------------32A}) and (\ref%
{Eq.-------------------32B}) take respectively the form}%
\begin{eqnarray}
&&\left. \frac{\partial L_{eff}}{\partial \chi (s)}=0,\right.  \label{EL-1}
\\
&&\left. \frac{\partial L_{eff}}{\partial u_{\mu }(s)}=0,\right.
\label{EL-2} \\
&&\left. \frac{d}{ds}\frac{\partial L_{eff}}{\partial \frac{dr^{\mu }(s)}{ds}%
}-\frac{\partial L_{eff}}{\partial r^{\mu }(s)}=0.\right.  \label{EL-3}
\end{eqnarray}%
\emph{These will be referred to as E-L equations in} \emph{standard
Lagrangian form.}\newline

\emph{T2}$_{2})$ \emph{The stress-energy tensor of the system $T_{\mu \nu }$
is uniquely determined in terms of }$L_{eff}.$ \emph{As a consequence, the
RR equation (\ref{Eq.---------------20B}) can also be written in
conservative form as}%
\begin{equation}
\overline{T}_{\mu \nu ,\nu }=0,  \label{conserv}
\end{equation}%
\emph{where }$\overline{T}_{\mu \nu }\equiv \overline{T}_{\mu \nu }^{\left(
M\right) }+\overline{T}_{\mu \nu }^{\left( EM\right) }$ \emph{is the
surface-averaged total stress energy tensor, obtained as the sum of the
corresponding tensors for the mass distribution and the EM field which
characterize the system. }

\bigskip

\emph{Proof} - \emph{T2}$_{1})$ The proof follows immediately by noting that
the Hamiltonian action (\ref{pointp}) defines a symmetric functional with
respect to local and non-local dependencies, i.e., such that%
\begin{equation}
S_{1}(r_{A},\left[ r_{B}\right] ,u,\chi )=S_{1}(r_{B},\left[ r_{A}\right]
,u,\chi ).
\end{equation}%
Because the E-L equations (\ref{EL-1})-(\ref{EL-3}) are written in terms of
local partial derivative differential operators, the effective Lagrangian $%
L_{eff}$ must be therefore distinguished from the corresponding variational
Lagrangian function $L_{1}$ which enters the Hamilton action and which
contains non-local contributions. These features imply the definition (\ref%
{extremal Lagrangian-0}), which manifestly satisfies the E-L equations in
standard form (\ref{EL-1})-(\ref{EL-3}).

\emph{T2}$_{2})$ The proof of this statement is straightforward, by first
recalling that the Lagrangian of the distributed mass is analogous to that
of a point mass particle. Moreover, the stress-energy tensor of the total EM
field $T_{\mu \nu }^{\left( EM\right) }$, to be defined in terms of $L_{eff}$
according to the standard definition (see for example Landau and Lifshitz
\cite{LL}) becomes%
\begin{equation}
T_{\mu \nu }^{\left( EM\right) }=T_{\mu \nu }^{\left( EM-ext\right) }+T_{\mu
\nu }^{\left( EM-self\right) }.
\end{equation}%
Then, given validity to the Maxwell equations, it follows that%
\begin{equation}
T_{\mu \nu ,\nu }^{\left( EM\right) }=F_{\mu \nu }j^{\nu }=\left[ F_{\mu \nu
}^{\left( ext\right) }+F_{\mu \nu }^{\left( self\right) }\right] j^{\nu }.
\end{equation}%
Gathering the mass and the field contributions, substituting the expressions
for $F_{\mu \nu }^{\left( ext\right) }$ and $F_{\mu \nu }^{\left(
self\right) }$ obtained in THM.1, and performing the integration over the
4-volume element finally proves that the equation (\ref{conserv}) actually
coincides with the extremal RR equation (\ref{Eq.---------------20B}).

\textbf{Q.E.D.}

\bigskip

The expression (\ref{conserv}) represents the conservative form of Eq.(\ref%
{Eq.---------------20B}), and hence - consistent with the surface
integration procedure here adopted - it holds for the \emph{surface-averaged}
EM external and self-fields $\overline{F}_{\mu \nu }^{(ext)}$ and $\overline{%
F}_{\mu \nu }^{\left( self\right) },$ defined respectively by Eqs.(\ref%
{EXTERNAL EM FIELD}) and (\ref{COVARIANT FORM}). It is important to remark
that the result holds both for finite-size and Lorentzian particles. On the
other hand, a local form of the conservative equation - analogous to Eq.(\ref%
{conserv}) - and holding for the local EM fields is in principle achievable
too. However, this last conclusion generally applies only to finite-size
particles with the same support for the mass and charge distributions, i.e.,
for which Eq.(\ref{SUPPORT}) holds.

\bigskip

\section{Short delay-time asymptotic approximation}

In this section we consider the asymptotic properties of the RR equation,
considering the customary approximation in the treatment of the problem,
which leads to the LAD equation (Dirac, 1938 \cite{Dirac1938}). This is the
power-series expansion of the retarded EM self-potential in terms of the
dimensionless parameter $\epsilon $\ $\equiv \frac{(s-s^{\prime })}{s},$ to
be assumed as infinitesimal (\emph{short delay-time ordering}), $s-s^{\prime
}$ denoting the proper-time difference between observation ($s$) and
emission ($s^{\prime }$). The same approach was also adopted by Nodvik \cite%
{Nodvik1964} in the case of flat space-time and by DeWitt and Brehme \cite%
{DeWitt1960} and Crowley and Nodvik \cite{Crowley-Nodvik1978} in their
covariant generalizations of the LAD equation valid in curved space-time. It
is immediate to show that the following result holds:\newline

\textbf{THM.3 - First-order, short delay-time asymptotic approximation}

\emph{Let us introduce the 4-vector }$G_{\mu }$ \emph{defined as }%
\begin{equation}
dsG_{\mu }=\frac{q}{c}\overline{F}_{\mu k}^{\left( self\right) }dr^{k},
\label{position}
\end{equation}%
\emph{and invoke the asymptotic ordering }
\begin{equation}
0<\epsilon \ll 1.  \label{asymptotic ordering}
\end{equation}%
\emph{\ Then:}

\emph{T3}$_{1})$\emph{\ Neglecting corrections of order }$\epsilon ^{N},$
\emph{with} $N\geq 1$ \emph{(first-order approximation})$,$\emph{\ the
following asymptotic approximation holds for }$G_{\mu }$%
\begin{equation}
\left. G_{\mu }\cong \left\{ -m_{oEM}c\frac{d}{ds}u_{\mu }+g_{\mu }\right\} %
\left[ 1+O(\epsilon )\right] \right. ,  \label{Eq.-beta}
\end{equation}%
\emph{where }$g_{\mu }$\emph{\ denotes the 4-vector}%
\begin{equation}
g_{\mu }=\frac{2}{3}\frac{q^{2}}{c}\left[ \frac{d^{2}}{ds^{2}}u_{\mu
}-u_{\mu }(s)u^{k}(s)\frac{d^{2}}{ds^{2}}u_{k}\right] ,  \label{Eq.-beta-0}
\end{equation}%
\emph{with}
\begin{equation}
m_{oEM}\equiv \frac{q^{2}}{c^{2}\sigma }\frac{1}{\left[ 1+\frac{(t-t^{\prime
})}{2}\frac{d}{ds}\frac{1}{\gamma }\right] ^{2}}  \label{Eq.-beta-00}
\end{equation}%
\emph{being the EM mass and }$\gamma \left( t(s)\right) \equiv 1/\sqrt{%
\left( 1-v^{2}(t(s)\right) /c^{2}}.$

\emph{T3}$_{2})$\emph{\ The point-charge limit of the RR equation \ (\ref%
{Eq.---------------20B})}$\ $\emph{does not exist.}

\emph{Proof} - \emph{T3}$_{1})$ The proof is straightforward and follows by
performing explicitly the perturbative expansion with respect to $\epsilon $%
. By dropping the terms which vanish in the limit $\epsilon \rightarrow 0,$
this yields Eq.(\ref{Eq.-beta}). The proof of \emph{T3}$_{2}),$ instead,
follows by noting that the limit obtained by letting
\begin{equation}
\sigma \rightarrow 0^{+}  \label{point-particle}
\end{equation}%
(\emph{point-charge limit}) is \emph{not defined,} since%
\begin{equation}
\lim_{\sigma \rightarrow 0^{+}}m_{oEM}=\infty .
\end{equation}%
\textbf{Q.E.D.}\newline

As basic consequences, in the first-order approximation the RR equation (\ref%
{Eq.---------------20B}) recovers the LAD equation. Moreover, in a similar
way, by introducing a suitable approximate reduction scheme, also the LL
equation (Landau and Lifschitz, 1951 \cite{LL}) can be immediately obtained.

\section{The fundamental existence and uniqueness theorem}

THMs.1 and 2 show that in the presence of RR the non-local Lagrangian
system\ $\left\{ \mathbf{x},L\right\} $ admits E-L equations [Eq.(\ref%
{Eq.---------------20B})] which are of \emph{delay differential type}. This
feature is not completely unexpected, since model equations of this type
have been proposed before for the RR problem (see for example \cite%
{caldirola}). In general, for a delay-type differential equation there is
nothing similar to the existence and uniqueness theorem holding for an
initial condition of the type%
\begin{equation}
\mathbf{x}(s_{o})=\mathbf{x}_{o}.  \label{Eq.-----------21C}
\end{equation}%
In fact, no finite set of initial data is generally enough to determine a
unique solution. The possibility of having, under suitable physical
assumptions, an existence and uniqueness theorem therefore plays a crucial
role in the proper formulation of the RR problem. In fact, for consistency
with the SR-CE axioms, and in particular with NPD, the existence of a
classical dynamical system (\ref{dynsys}) must be warranted. The result can
be obtained by requiring that there exists an initial time $s_{o}$ before
which for all $s<s_{o}$ the particle motion is inertial (see also the
related discussion in Ref.\cite{caldirola}). The assumption has also been
invoked to define the particle mass and charge distributions (see Section
3). In view of THM.1 this happens if the external EM force vanishes
identically for all $s<s_{o}$ and is (smoothly) \textquotedblleft turned
on\textquotedblright\ at $s=s_{o}$. In this regard, we here point out the
following theorem:\newline

\textbf{THM.4 - The fundamental theorem for the RR equation}

\emph{Given validity of THM.1, let us assume that:}

\begin{enumerate}
\item \emph{REQUIREMENT \#1: at time }$t_{o}$ \emph{the initial condition (%
\ref{Eq.-----------21C}) holds;}

\item \emph{REQUIREMENT \#2: the external force }$\overline{F}_{\mu \nu
}^{(ext)}(r,s)$ \emph{is of the form }$\overline{F}_{\mu \nu
}^{(ext)}(r,s)=\Theta (s-s_{o})\overline{F}_{1\mu \nu }^{(ext)}(r)),$ \emph{%
i.e., }$\overline{F}_{\mu \nu }^{(ext)}$ \emph{is \textquotedblleft turned
on\textquotedblright\ at the proper time }$s=s_{o}$. \emph{In particular we
shall take $\overline{F}_{\mu \nu }^{(ext)}(r,s)$ to be a smooth function of
$s$, of class $C^{k}\left( M^{4}\times I\right) $, with }$k\geq 1$\emph{;}

\item \emph{REQUIREMENT \#3: more generally, let us require that for an
arbitrary initial state }$\mathbf{x}(s_{1})=\mathbf{x}_{1}$\emph{\ }$\in
\Gamma $\emph{\ there always exists }$\left\{ \mathbf{x}(s_{o})=\mathbf{x}%
_{o},s_{o}\right\} \in \Gamma \times I,$\emph{\ with }$s_{o}=s_{1}-s_{ret},$%
\emph{\ such that at time }$s_{o},$\emph{\ }$\mathbf{x}(s_{o})$\emph{\ is
inertial, i.e., before s}$_{o}$\emph{\ the external force }$\overline{F}%
_{\mu \nu }^{(ext)}$ \emph{vanishes identically, so that the dynamics is of
the form provided by Eqs.(\ref{INERTIAL-1})-(\ref{INERTIAL-2}). }
\end{enumerate}

\emph{It then follows that the solution of the initial-value problem (\ref%
{Eq.---------------20B})-(\ref{Eq.-----------21C}), subject to REQUIREMENTS
\#1-\#3, exists at least locally in a subset }$I\equiv \left[ -\infty ,s_{0}%
\right] \cup \left[ s_{0},s_{n}\right] \subseteq
\mathbb{R}
$ \emph{with }$\left[ s_{0},s_{n}\right] $\emph{\ a bounded interval,} \emph{%
and is unique (\textit{fundamental theorem}). }

\emph{Proof} - Eq.(\ref{Eq.---------------20B}) can be cast in the form of a
delay-differential equation, i.e.,%
\begin{equation}
\frac{d\mathbf{x}(s)}{ds}=\mathbf{X(x}(s),\mathbf{x}(s-s_{ret}),s),
\label{Eq.------------------24}
\end{equation}%
subject to the initial condition
\begin{equation}
\mathbf{x}(s_{o})=\mathbf{x}_{o}.  \label{Eq.-------------------24B-}
\end{equation}%
Here $\mathbf{x}(s)$ and $\mathbf{x}(s-s_{ret})$ denote respectively the
\textquotedblleft instantaneous\textquotedblright\ and \textquotedblleft
retarded\textquotedblright\ states $\mathbf{x}(s)$ and $\mathbf{x}%
(s-s_{ret}),$ while $\mathbf{X(x}(s),\mathbf{x}(s-s_{ret}),s)$ is a suitable
$C^{2}$ real vector field depending smoothly on both of them. The proof of
local existence and uniqueness for Eq.(\ref{Eq.------------------24}), with
the initial conditions (\ref{Eq.-------------------24B-}) and the
Requirements \#1-\#3, requires a generalization of \textit{the fundamental
theorem }holding for ordinary differential equations\ (in which the vector
field $\mathbf{X}$ depends only on the local state $\mathbf{x}(s)$).

Let us first consider the case in which the solution $\mathbf{x}(s)$ of the
initial-value problem\ (\ref{Eq.------------------24}) and (\ref%
{Eq.-------------------24B-}) is defined in the half-axis $\left[ -\infty
,s_{o}\right] :$ by assumption this solution exists, is unique and is that
of inertial motion [see Eqs.(\ref{INERTIAL-1})-(\ref{INERTIAL-2})].

Next, let us consider the proper time interval $I_{o,1}\equiv \left[
s_{o},s_{1}\equiv s_{o}+s_{ret}\right] .$ Thanks to the Requirement \#3, by
assumption in $I_{o,1}$ the particle is subject only to the action of the
external force (produced by $A_{\mu }^{(ext)}$), since $\overline{F}_{\mu
\nu }^{\left( self\right) }$ vanishes by definition if $s<s_{o}+s_{ret} $.
Hence, in the same time interval the solution exists and is unique because
the differential equation (\ref{Eq.------------------24}) is of the form%
\begin{equation}
\frac{d\mathbf{x}(s)}{ds}=\mathbf{X}^{ext}\mathbf{(x}(s),s),  \label{form-1}
\end{equation}%
with $\mathbf{X}^{ext}\mathbf{(x}(s),s)$ being, by assumption, a smooth
vector field (see THM.1). Eq.(\ref{form-1}) is manifestly a local ODE for
which the fundamental theorem (for local ODEs) holds. Hence, existence and
uniqueness is warranted also in $I_{o,1}$.

Finally, let us consider the sequence of proper time intervals $%
I_{k,k+1}\equiv \left[ s_{k},s_{k+1}=s_{k}+s_{ret}\right] ,$ for the integer
$k=1,2,3...n$, where $n$ $\geq 2.$ In this case, for any proper time $s\in
I_{k,k+1},$ the advanced-time solution $\mathbf{x}(s-s_{ret})$ appearing in
the vector field $\mathbf{X\equiv X(x}(s),\mathbf{x}(s-s_{ret}),s)$ can be
considered as a \emph{prescribed function of} $s,$ determined in the
previous time interval $I_{k,k-1}.$ Therefore, $\mathbf{X}$ is necessarily
of the form $\mathbf{X\equiv }\widehat{\mathbf{X}}\mathbf{(x}(s),s)$, so
that for $s>s_{1},$ Eq.(\ref{Eq.------------------24}) can be viewed again
as a local ODE. We conclude that, thanks to the fundamental theorem holding
for local ODEs, the local existence (in a suitable bounded proper time
interval $I\equiv \lbrack s_{1},s_{n}]$) and uniqueness of solutions of the
problem (\ref{Eq.------------------24})-(\ref{Eq.-------------------24B-})
is assured under the Requirements \#1-\#3. This proves the statement.

\textbf{Q.E.D.}

\section{Conclusions}

In this paper we have shown that the RR problem originally posed by Lorentz
for classical non-rotating finite-size and Lorentzian particles can exactly
be solved analytically within the SR setting.

For these particles, the resulting relativistic dynamics in the presence of
the RR force, i.e., the \emph{classical RR equation}, has been found
analytically by taking into account the exact covariant form of the EM self
4-potential. In particular, this has been uniquely determined consistently
with the basic principles of classical electrodynamics and special
relativity. In addition, the RR equation has been proved to be\emph{\
variational }in the functional class of synchronous variations (\ref%
{FUNCTIONAL CLASS}) with respect to the Hamilton variational principle,
defined in terms of a non-local variational Lagrangian function. The same
equation has been shown: 1) to admit the standard Lagrangian form in terms
of the non-local effective Lagrangian $L_{eff}$; 2) to admit a conservative
form; 3) to recover the usual \emph{asymptotic} LAD and LL equations in the
first-order short delay-time approximation; 4) not to admit the point-charge
limit. From the mathematical point of view, the RR equation is a delay-type
second order ODE, which fulfills \emph{GIP }in the sense of THM.1\emph{,
relativistic covariance and MLC}. As a consequence, provided suitable
physical requirements are imposed, \emph{the initial-value problem for the
RR equation is well-posed}, defining the classical dynamical system required
by NDP.

\acknowledgments This work was developed in the framework of the research
projects of the Consortium for Magnetofluid Dynamics (University of Trieste,
Italy): \textit{Fundamentals and applications of relativistic Hydrodynamics
and Magnetohydrodynamics }(International School for Advanced Studies
(SISSA), Trieste, Italy) and\textit{\ Magnetohydrodynamics in curved space:
theory and applications }(Department of Mathematics and Informatics,
University of Trieste, Italy).

\bigskip

\section{Appendix A: variational calculations}

Here we report the proof of identities (\ref{Eq.---A.3})-(\ref{A_nu_k}) in
THM.1. Let us first notice that
\begin{equation}
d\left[ \int_{1}^{2}dr^{\prime \nu }\delta (\widetilde{R}^{\alpha }%
\widetilde{R}_{\alpha }-\sigma ^{2})\right] =dr^{k}\int_{-\infty }^{\infty
}ds^{\prime }u^{\nu }(s^{\prime })\frac{\partial }{\partial r^{k}}\left[
\delta (\widetilde{R}^{\alpha }\widetilde{R}_{\alpha }-\sigma ^{2})\right] .
\end{equation}%
Hence the variations $\delta A$ and $\delta B$\ given in Eqs.(\ref{EQ.---A.5}%
) are respectively
\begin{equation}
\delta A=-\frac{4q^{2}}{c}\eta _{\mu \nu }\int_{1}^{2}\delta r^{\mu
}dr^{k}\int_{-\infty }^{\infty }ds^{\prime }u^{\nu }(s^{\prime })\frac{%
\partial }{\partial r^{k}}\left[ \delta (\widetilde{R}^{\alpha }\widetilde{R}%
_{\alpha }-\sigma ^{2})\right] ,  \label{za}
\end{equation}%
while%
\begin{equation}
\delta B=\frac{4q^{2}}{c}\eta _{\alpha \beta }\int_{1}^{2}dr^{\alpha }\delta
r^{\mu }\int_{-\infty }^{\infty }ds^{\prime }u^{\beta }(s^{\prime })\frac{%
\partial }{\partial r^{\mu }}\delta (\widetilde{R}^{k}\widetilde{R}%
_{k}-\sigma ^{2}).  \label{zb}
\end{equation}%
Let us now evaluate the partial derivative $\frac{\partial }{\partial r^{k}}%
\delta (\widetilde{R}^{\alpha }\widetilde{R}_{\alpha }-\sigma ^{2}).$
Invoking the chain rule, this becomes%
\begin{equation}
\frac{\partial }{\partial r^{k}}\delta (\widetilde{R}^{\alpha }\widetilde{R}%
_{\alpha }-\sigma ^{2})=\frac{\partial (\widetilde{R}^{\alpha }\widetilde{R}%
_{\alpha })}{\partial r^{k}}\frac{d\delta (\widetilde{R}^{\alpha }\widetilde{%
R}_{\alpha }-\sigma ^{2})}{d(\widetilde{R}^{\alpha }\widetilde{R}_{\alpha })}%
=\frac{d\delta (\widetilde{R}^{\alpha }\widetilde{R}_{\alpha }-\sigma ^{2})}{%
ds^{\prime }}\frac{2\widetilde{R}_{k}}{\frac{d(\widetilde{R}^{\alpha }%
\widetilde{R}_{\alpha })}{ds^{\prime }}},
\end{equation}%
and so%
\begin{equation}
\frac{\partial }{\partial r^{k}}\delta (\widetilde{R}^{\alpha }\widetilde{R}%
_{\alpha }-\sigma ^{2})=-\frac{\widetilde{R}_{k}}{\widetilde{R}^{\alpha
}u_{\alpha }(s^{\prime })}\frac{d}{ds^{\prime }}\left\{ \frac{\delta
(s-s^{\prime }-s_{ret})}{2\left\vert \widetilde{R}^{\alpha }u_{\alpha
}(s^{\prime })\right\vert }\right\} .  \label{az}
\end{equation}%
It follows that%
\begin{eqnarray}
\frac{\partial }{\partial r^{k}}\delta (\widetilde{R}^{\alpha }\widetilde{R}%
_{\alpha }-\sigma ^{2}) &=&-\frac{\widetilde{R}_{k}}{c^{2}\left[
(t-t^{\prime })-\frac{1}{c^{2}}\frac{d\mathbf{r}^{\prime }}{dt^{\prime }}%
\cdot (\mathbf{r-r}^{\prime })\right] }\times  \notag \\
&&\times \frac{d}{dt^{\prime }}\left\{ \frac{\delta (t-t^{\prime }-t_{ret})}{%
2c^{2}\gamma \left( t^{\prime }\right) \left\vert (t-t^{\prime })-\frac{1}{%
c^{2}}\frac{d\mathbf{r}(t^{\prime })}{dt^{\prime }}\cdot (\mathbf{r-r}%
^{\prime })\right\vert }\right\} ,  \label{z}
\end{eqnarray}%
where $\mathbf{r}^{\prime }\equiv \mathbf{r}(t^{\prime }),$ $t\equiv t(s)$
and $t^{\prime }\equiv t(s^{\prime }).$ Substituting Eq.(\ref{z}) into Eqs.(%
\ref{za}) and (\ref{zb}), and then directly integrating, it follows
immediately that $\delta A$ and $\delta B$ have the form (\ref{EQ.---A.5}).

\end{document}